\documentclass[%
reprint,
superscriptaddress,
 amsmath,amssymb,
 aps,
twocolumn,
floatfix,
]{revtex4}

\usepackage{graphicx}% Include figure files
\usepackage{dcolumn}% Align table columns on decimal point
\usepackage{bm}% bold math
\usepackage{xspace}
\usepackage{color}
\usepackage[]{hyperref}
  \hypersetup{      
  unicode=true,         
  pdftoolbar=true,     
  pdfmenubar=true,     
  pdffitwindow=true,    
  pdfstartview={FitH},    
  pdfsubject={Sterilenus@DUNE},  
  pdfnewwindow=true,     
  pdfcreator={RevTeX},
  colorlinks=true,     
  linkcolor=red,   
  citecolor=blue,    
  filecolor=black,   
  urlcolor=blue,           
  }
\usepackage{placeins}
\usepackage{diagbox}
\newcommand{\gev}{\ensuremath{\,\text{GeV}}\xspace}

\mathchardef\-="2D

\begin{document}

\preprint{APS/123-QED}

\title{Estimating the Uncertainty of Cosmological First Order Phase Transitions with Numerical Simulations of Bubble Nucleation}

\affiliation{
CAS Key Laboratory of Theoretical Physics, Institute of Theoretical Physics, Chinese Academy of Sciences, Beijing 100190, P. R. China}

\author{Huai-Ke Guo}
\email[]{guohuaike@ucas.ac.cn}
\affiliation{ International Centre for Theoretical Physics Asia-Pacific, University of Chinese Academy of Sciences, Beijing 100190, P. R. China}

\author{Song Li}
\email[Corresponding author, ]{lisong@itp.ac.cn}
\affiliation{
CAS Key Laboratory of Theoretical Physics, Institute of Theoretical Physics, Chinese Academy of Sciences, Beijing 100190, P. R. China}
\affiliation{School of Physical Sciences, University of Chinese Academy of Sciences, Beijing 100049, P. R. China}

\author{Yang Xiao}
\email[Corresponding author, ]{xiaoyang@itp.ac.cn}
\affiliation{
CAS Key Laboratory of Theoretical Physics, Institute of Theoretical Physics, Chinese Academy of Sciences, Beijing 100190, P. R. China}
\affiliation{School of Physical Sciences, University of Chinese Academy of Sciences, Beijing 100049, P. R. China}

\author{Jin Min Yang}
\email[]{jmyang@itp.ac.cn}
\affiliation{
CAS Key Laboratory of Theoretical Physics, Institute of Theoretical Physics, Chinese Academy of Sciences, Beijing 100190, P. R. China}
\affiliation{School of Physical Sciences, University of Chinese Academy of Sciences, Beijing 100049, P. R. China}

\author{Yang Zhang}
\email[]{zhangyangphy@zzu.edu.cn}
\affiliation{School of Physics, Zhengzhou University, Zhengzhou 450000, P. R. China}
\affiliation{Institute of Physics, Henan Academy of Sciences, Zhengzhou 450046, P. R. China}

\date{\today}% It is always \today, today,

\begin{abstract}

% We generated a large number of vacuum bubbles based on different bubble nucleation rates and simply simulated the kinematics of the bubbles. 
In order to study the validity of analytical formulas used in the calculation of characteristic physical quantities related to vacuum bubbles, we conduct several numerical simulations of bubble kinematics in the context of cosmological first-order phase transitions to determine potentially existing systematic uncertainties. 
% In order to obtain the suitable simulated volume that could reproduce the real phase transition of the universe as faithfully as possible, we conduct several numerical simulations of bubble kinematics in the context of cosmological first-order phase transitions. 
By comparing with the analytical results, we obtain the following observations: (1) The simulated false vacuum fraction will approach the theoretical one with increasing simulated volume. When the side length of the cubic simulation volume becomes larger than $14.5\beta_{\rm th}^{-1}$, the simulated results do not change significantly; (2) The theoretical expected total number of bubbles do not agree with the simulated ones, which may be caused by the inconsistent use of the false vacuum fraction formula; (3)  The different nucleation rate prefactors do not affect the bubble kinetics much; (4)  The lifetime distribution in the sound shell model does not obey an exponential distribution, in such a way as to cause a suppression in the gravitational wave spectra.

\end{abstract}

\maketitle

\section{Introduction} 
With the first direct detection of gravitational waves induced by the merger of binary black holes~\cite{abbott2016observation}, the era of gravitational wave observations has arrived. One of the next important goals is to search for stochastic gravitational waves from the cosmic first-order phase transition and probe the possible new physics beyond the standard model~\cite{weir2018gravitational, mazumdar2018cosmic, bertone2020gravitational}. To the end, LIGO and Virgo have already taken the first step~\cite{abbott2017upper, scientific2019search} and many space-based detectors have been proposed, such as the Laser Interferometer Space Antenna (LISA)~\cite{amaro2017laser}, Taiji~\cite{gong2015descope} and Tianqin~\cite{luo2016tianqin}. These detectors are expected to be completed in a decade or so and start to search for low-frequency gravitational waves induced at the electroweak scale~\cite{Athron:2023xlk,han2021dark,guo2021phase,yang2023gravitational,Xiao:2023dbb,Balazs:2023kuk,vaskonen2017electroweak,beniwal2019gravitational,kang2018strong,chala2018signals, alves2020di, alves2019collider, chao2017gravitational}. 

In terms of phenomenological analysis, it is well known that the source of phase transition gravitational wave comes from three parts: bubble collision, sound wave, and turbulence. The contribution of bubble collision to observable gravitational waves can be well described by the envelope approximation~\cite{kosowsky1992gravitational, kosowsky1993gravitational} and the corresponding analytical formula has already been derived~\cite{jinno2017gravitational}. Furthermore, lattice simulation results can provide more information than the envelope approximation and have yielded the standard spectral formulas that are widely used~\cite{hindmarsh2015numerical,hindmarsh2017shape}. The lattice results show that the contribution of sound waves is much larger than the collisions~\cite{hindmarsh2014gravitational, hindmarsh2015numerical, hindmarsh2017shape}. To describe it, many simplified models have been developed, such as the sound shell model~\cite{hindmarsh2018sound, hindmarsh2019gravitational} and the bulk flow model \cite{jinno2019gravitational, konstandin2018gravitational} as well as their extensions~\cite{cai2023hydrodynamic}, in which the power laws can be accurately reproduced. In contrast to the first two parts, the early understanding of turbulence was almost entirely derived from theoretical calculations~\cite{kosowsky2002gravitational,dolgov2002relic, Caprini_2009}, whereas recently numerical simulations have started to be performed~\cite{Yang:2021uid, Di:2020kbw}.

To keep up with advances in experiments,  there is increasing attention being paid to the study of uncertainties in the calculations mentioned above, such as the uncertainties in the effective potential~\cite{croon2021theoretical,Athron:2022jyi} and in the calculation of bounce actions~\cite{andreassen2017precision,dunne2005beyond, Ivanov:2022osf} and nucleation rate~\cite{ekstedt2022higher}. In this work, we will discuss the uncertainties involved in estimating the pre-factor in the nucleation rate and the finite volume effect on simulation.

The uncertainty of the pre-factor arises from the determinant related to the potential, which is model-dependent and difficult to obtain with high precision. Recently, the package \texttt{BubbleDet} is developed to compute this functional determinant quickly and easily~\cite{Ekstedt:2023sqc}. However, for convenience, it is typically approximately estimated using one of three approaches: (1) taking the fourth power of the critical temperature~\cite{hindmarsh2019gravitational, guth1981cosmological}, (2) taking the fourth power of temperature~\cite{enqvist1992nucleation, guo2021phase, han2021dark, alves2019collider,chao2017gravitational}, and (3) taking the fourth power of temperature while incorporating the bounce action~\cite{athron2023supercool, cai2017gravitational}.
The use of different pre-factors may lead to different rates of nucleation, resulting in different kinematic processes in both analytical calculations and numerical simulations. 

Due to the extensive computational resources necessitated for lattice calculations, current numerical simulations of gravitational waves are constrained to relatively modest volumes.
Although almost all simulations adopt characteristic physical parameters, such as critical temperature, as a measure for length, it remains uncertain to what degree these simulations accurately capture actual cosmological phase transition processes.
Furthermore, in the simulations vacuum bubbles is generated randomly based on an approximate nucleation rate given by $\Gamma_{\rm approx} = p_0 e^{-\beta (t-t_f)}$, where $p_0$ and $\beta$ must be assigned appropriate values to generate a sufficient number of bubbles in a limited volume.
For instance, in the simulation of Ref.\cite{hindmarsh2014gravitational}, $p_0 = 0.01$, $\beta = 0.0125 T_{\rm c}$ were adopted. 
However, for electroweak phase transitions, $\beta$ typically ranges from $\mathcal{O}(10^{-11}) \sim \mathcal{O}(10^{-13})\gev$, and $p_0 \propto T^4$. 
While we presume that the values of these parameters will not significantly influence the electroweak phase transition process, the errors stemming from the finite simulation volume are nonetheless crucial and may serve as a valuable reference for future detections. 
Given the vastness of the universe and the availability of expressions for the false vacuum fraction in the infinite volume limit, it is intriguing to leverage the discrepancy between theoretical formulations and simulation outcomes to assess whether the corresponding simulated volume is sufficiently expansive.

Both the nucleation rate and the false vacuum fraction are essential factors in describing bubble kinematics. In the sound shell model, many statistical measures related to gravitational waves are dependent on these factors, such as the bubble mean separation and the bubble lifetime distribution. Therefore, any variation in these factors may potentially affect the predicted gravitational wave spectra. Adopting simple simulation to confirm if those theoretical expression can well describe the real process and find out how much difference on the gravitational wave spectra can arising is also a meaningful investigation.

The aim of this work is to conduct a bubble simulation based on a given nucleation rates and compare the simulated values with the analytical results for the false vacuum fraction, the total number of bubbles, and the bubble lifetime distribution to obtain insights into the uncertainties. The paper is structured as follows: in section II, we present the theoretical framework for the relevant quantities; section III describes the simulation details; in section IV, we compare the simulation results with the theoretical predictions; and finally, we conclude in section V.

\section{Physical quantities in bubble kinematics}

During the early universe, bubbles of true vacuum formed within the false vacuum environment.
%The true vacuum bubble is nucleated in the universe filled with false vacuum. 
The nucleation rate, i.e., the number of bubbles nucleated per unit time per unit volume, can be roughly estimated as~\cite{linde1983decay}
\begin{equation} \label{eq: Gama}
    \Gamma \sim A e^{-S_3/T},
\end{equation}
where $A$ is a pre-factor and $S_{3}$ is the three-dimensional Euclidean action given by
\begin{equation}
    S_{3} =  4 \pi \int^{+\infty}_{0}r^2{\rm d}r~\left[\frac{1}{2}\left(\frac{\partial \phi}{\partial r}\right)^{2} + V(\phi,T)\right].
\end{equation}
The bubble configuration $\phi(r)$ in the integral is fixed from the corresponding Euclidean equation of motion
\begin{equation}
    \frac{{\rm d}^2 \phi}{{\rm d} r^2} + \frac{2}{r}\frac{{\rm d} \phi}{{\rm d}r} = \frac{\partial V(\phi,T)}{\partial \phi},
\end{equation}
subjected to the boundary conditions (see Refs.~\cite{rubakov2009classical,linde1983decay} for details)
\begin{equation}
    \lim \limits_{r \to \infty} \phi(r) = 0, ~~ {\rm d} \phi/{\rm d} r|_{r=0}=0.
\end{equation} 

The pre-factor is proportional to the fourth power of temperature on dimensional grounds, and more precisely its specific value is given by~\cite{linde1983decay}
\begin{equation} \label{eq: det}
    A = T\left(\frac{S_3}{2 \pi T}\right)^{\frac{3}{2}} \frac{{\rm det}'\left[-\Delta + V''(\phi,T)\right]}{{\rm det}\left[-\Delta + V''(0,T)\right]},
\end{equation}
where $\rm{det}'$ implies that three zero eigenvalues of the operator $-\Delta + V''(\phi,T)$ are to be omitted when computing the determinant. Its precise determination requires integrating out fluctuations around the bounce solution, as demonstrated in Refs.\cite{andreassen2017precision,dunne2005beyond}. 
For simplification, three approximate estimations have been proposed and are commonly used:
\begin{enumerate}
    \item $A \approx T_{\rm c}^4$, where $T_{\rm c}$ is the critical temperature,
    \item $A \approx T^4$ ,
    \item $A \approx T^4 (\frac{S_3}{2 \pi T})^{\frac{3}{2}}$.
\end{enumerate}
In the following, we refer them as pre-factor1, pre-factor2 and pre-factor3, respectively. These modes will yield different nucleation rates and, consequently, may affect the bubble kinematics later on.

After nucleation, the bubbles may expand and merge until the entire universe is filled with true vacuum. To account for the remained fraction of space in the false vacuum, we follow Guth and Weinberg's procedure~\cite{guth1981cosmological} and consider a space containing randomly placed spheres (including overlapped and nested spheres). We then calculate the probability that a given point is not contained within any existing sphere. 

Denoting $\rho(V) dV$ as the density of spheres with volume between $V$ and $V+dV$, $V_{\rm s}$ as the volume of the entire space, and $g(V_1,V_2)$ as the probability that a given point is not contained in any existing sphere with volume between $V_1$ and $V_2$, we can derive the following equation,
\begin{equation} \label{eq: Guth}
\begin{aligned}
    g(V_1,V_2+dV_2) =& g(V_1, V_2) g(V_2, V_2+dV_2)\\
    =& g(V_1, V_2) \frac{V_{\rm s}-V_{\rm s} \rho(V_2) dV_2V_2}{V_{\rm s}}\\
    =& g(V_1, V_2) (1 - \rho(V_2)dV_2V_2),       
\end{aligned}
\end{equation}
which then leads to a differential equation for $g$:
\begin{equation}
    \frac{d g(V_1, V_2)}{d V_2} = -\rho(V_2)g(V_1,V_2)V_2.
\end{equation}
Note that, to have a physical $g(V_2, V_2+dV_2)$, $V_{\rm s}$ must exceed than $V_{\rm s} \rho(V_2) dV_2V_2$. This implies that $V_{\rm s}$ should be considered as infinity to guarantee that this inequality holds for all possible $\rho$.
The solution of this equation is 
\begin{equation}
    g(V_1,V_2) = {\rm{exp}}\left[-\int_{V_1}^{V_2}dV \rho(V)V\right].
\end{equation}
In particular, the probability that the given point is not contained within any existing sphere is
\begin{equation}
    g(0, \infty) = {\rm{exp}}\left[-\int_{0}^{\infty}dV \rho(V)V\right].
\end{equation}

Returning to the picture of the first-order transition and taking into account the fact that the bubbles forming at the same time have the same volume, we can substitute $\rho(V) dV$ with $\Gamma(t) dt$. Therefore, in a flat spacetime, the fraction of false vacuum can be approximated as 
\begin{equation} \label{eq: h(t)}
  h_{\rm th}(t) \equiv g(t_{\rm c}, t) = {\rm{exp}}\left[-\int^t _{t_{\rm c}}\Gamma(t')V(t',t)dt'\right], 
\end{equation}
where $V(t',t) = \frac{4\pi}{3} [\int ^t _{t'} v_{\rm w}(\tau) d\tau]^3$ and $v_{\rm w}$ is the bubble wall velocity.
The same equation can be derived from different perspectives, as demonstrated in Ref.\cite{Ajmi:2022nmq, Athron:2023xlk}. 

If we expand the action in the vicinity of a reference time $t_0$, we can simplify Eq.~(\ref{eq: Gama}) as 
\begin{equation} \label{eq: approx nucleation rate}
    \Gamma = \Gamma_0 e^{\beta(t-t_0)}
\end{equation}
where $\beta = \left. -d {\rm{ln}}\Gamma/dt \right|_{t=t_0}$. Moreover, assuming a constant pre-factor, Eq.~(\ref{eq: h(t)}) can also be further simplified to~\cite{enqvist1992nucleation}:
\begin{equation} \label{eq: simpifiled h}
    h_{\rm approx}(t) = {\rm{exp}}\left[-e^{\beta (t-t_0)}\right].
\end{equation}

Obtaining a closed-form solution for the false vacuum fraction in a finite volume is challenging, necessitating the employment of Monte Carlo simulations. A more accurate and efficient, albeit complex, alternative approach involves the utilization of Voronoi diagrams in Laguerre geometry~\cite{imai1985voronoi}. While Eq.~(\ref{eq: h(t)}) precisely depicts the phase transition scenario in an infinite volume, the results of finite-size simulations may exhibit discrepancies. It is intuitive to anticipate that these discrepancies will diminish as the simulated volume increases. Consequently, we can quantify these discrepancies to determine the optimal simulated volume that most accurately reflects the real universe.

We now have two key ingredients, i.e., the false vaccum fraction and the nucleation rate. It is worth noting that bubbles can only be generated in the false vacuum region. As a result, the total number of nucleated bubbles before time $t$ within the volume $V_{\rm s}$ can be expressed as~\cite{enqvist1992nucleation}
\begin{equation} \label{eq: total number}
    N_{\text{tot}}(t) = \int_{t_{\rm c}} ^{t} V_{\rm s} \Gamma(t') h(t') dt',
\end{equation}
where $t_{\rm c}$ is the evolution time corresponding to the critical temperature $T_{\rm c}$, $h$ is the false vacuum fraction. The mean separation between bubbles can be defined as 
\begin{equation}
    R^{*} = \left(\frac{V_{\rm s}}{N_{\rm tot}}\right)^{\frac{1}{3}}.
\end{equation}

\begin{figure}[thbp!]
\centering 
\includegraphics[width=0.49\textwidth]{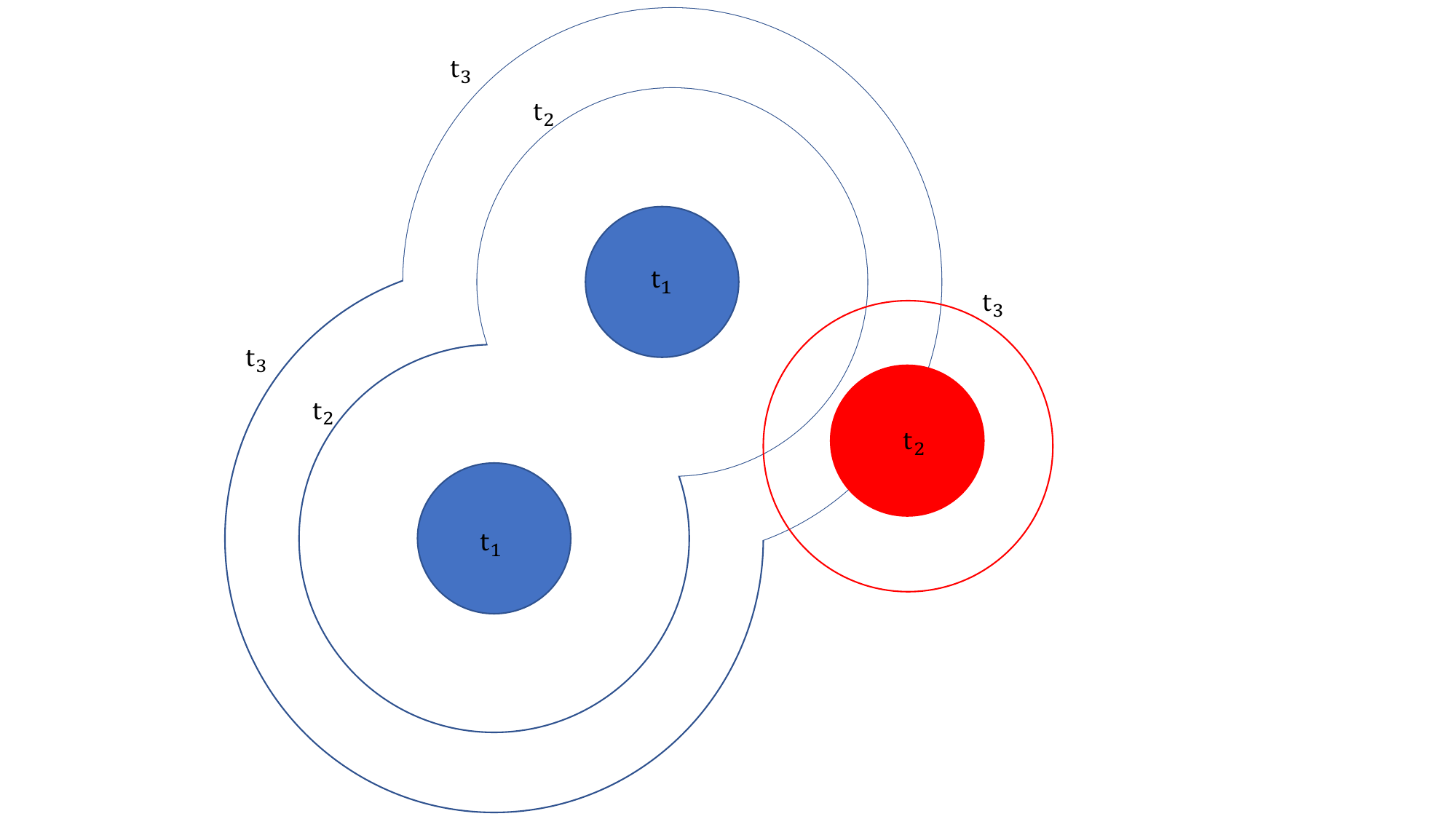}
\caption{The schematic diagram illustrating a scenario for calculating the lifetime of the red bubble. At time $t_1$, two blue bubbles are generated randomly. The two blue bubbles merge at $t_2$, and contact the nucleation site of the $t_2$ birthed red bubble at $t_3$. In our definition, the red bubble is destroyed at $t_3$ and its lifetime is $t_3 - t_2$.}
\label{fig:1}
\end{figure}

Another useful statistical measure is the bubble lifetime distribution, which characterizes the lifetimes for all the bubbles that were formed and subsequently destroyed during the phase transition process~\cite{hindmarsh2019gravitational, hijazi2022numerical}. A bubble is considered as being destroyed when the nucleation site is first occupied by the true vacuum space, as illustrated in Fig.~\ref{fig:1}. To analyze the bubble lifetime distribution, we first define the unbroken bubble wall area per unit volume $S$ as~\cite{hindmarsh2019gravitational, guo2021phase} 
\begin{equation}
   \frac{dV_{\rm false}}{V_{\rm s}} =  -SdR = -Sv_{\rm w}dt.
\end{equation}
The number of bubbles with a radius of $R$ at the time of their destruction, which occurs after a time $\bar{t} = R/v_{\rm w}$, is given by the following expression (see Fig.\ref{fig:1} for more details):
\begin{equation}
    d^2n = S(t' + R/v_{\rm w})\Gamma(t')dRdt'.
\end{equation}
By integrating this expression, we can obtain the bubble size distribution,
\begin{equation}
    \frac{dn}{dR} = \int_{t_{\rm c}} ^{\infty} S(t' + R/v_{\rm w}) \Gamma(t') dt'.
\end{equation}
The fraction of bubbles with lifetime in the range $\left[\bar{t}, \bar{t} + d\bar{t} \right]$ is 
\begin{equation}
    f = \frac{dn}{dR} \frac{dR}{d\bar{t}} =  v_{\rm w} \frac{dn}{dR}.
\end{equation}
Adopting Eq.(\ref{eq: simpifiled h}), we can express the normalized bubble lifetime distribution $\nu$ simply as~ \cite{hindmarsh2019gravitational}
\begin{equation}\label{eq: lifetime}
    \nu(\beta \bar{t}) = e^{-\beta \bar{t}}. 
\end{equation}

Assuming that the bubble will attain a steady velocity as a result of friction, the velocity profile $v_{\rm plasma}$ and enthalpy profile $w_{\rm plasma}$ of the plasma surrounding a single bubble can be calculated by using the hydrodynamics within a simplified bag model~\cite{espinosa2010energy}. Those self-similar profiles become the initial condition for freely propagating, after some forced propagating period~\cite{Cai:2023guc}. In the sound shell model, the total velocity profile is assumed to be a superposition of these velocity perturbations~\cite{hindmarsh2018sound, hindmarsh2019gravitational}
\begin{equation}
    v_{\rm plasma}^{\rm tot} = \sum_{i=1}^{\rm{N_{tot}}} v_{\rm plasma}^{i} .
\end{equation}
Then the velocity power spectrum can be derived
\begin{equation}  \label{eq: Pv}
    \mathcal{P}_{v}(q) = \frac{2}{2\pi^2(\beta R^{*})^3} (\frac{q}{\beta})^3 \int \nu(\tilde{T}) \tilde{T}^{6} \left| A(\tilde{T}q/\beta) \right|^{2} d\tilde{T},
\end{equation}
where
\begin{equation}
\begin{aligned}
    A(z) &= 0.5 (f'(z) + ic_{s}l(z)), \\
    f(z) &= \frac{4\pi}{z} \int v_{\rm plasma}(\xi) \sin (z \xi) d\xi, \\
    l(z) &= \frac{4 \pi}{z} \int \frac{e_{\rm plasma}(\xi) - \bar{e}}{\bar{w}} \xi \sin(z \xi) d\xi, 
\end{aligned}
\end{equation}
and $\bar{e}, \bar{w}$ are the mean energy and enthalpy densities, respectively. The gravitational wave power spectrum before redshift $\mathcal{P}_{\text{GW}}$ is defined as the contribution to the density fraction in gravitational waves per logarithmic wave number interval. With the help of $ \mathcal{P}_{v}$, the gravitational wave power spectrum generated by the stationary velocity power spectrum with a lifetime $\tau_{v}$ can be expressed as 
\begin{equation}
     \mathcal{P}_{\text{GW}} = 3(\Gamma_{w} \bar{U}_{f}^2)^2 (H \tau_{v})(H L_{f})\frac{(qL_{f})^3}{2\pi^2} \tilde{P}(qR^*),
\end{equation}
where $\Gamma_w$ is the ratio of enthalpy to energy, $\bar{U}_{f}$ is the RMS fluid velocity, $L_{f}$ is the length scale in the velocity field and 
\begin{eqnarray}
    &&\tilde{P}(y) = \frac{1}{4 \pi y c_{\rm s}} \left(\frac{1 - c_{s}^2}{c_{s}^2}\right)^2 \nonumber \\
    &&\times\int^{z_{+}}_{z{-}} 
      \frac{dz}{z} \frac{(z-z_{+})^2(z-z_{-})^2}{z_{+}+z_{-}-z}\bar{P}_{v}(z)\bar{P}_{v}(z_{+} + z_{-} - z) ,\nonumber \\
\end{eqnarray}
with $\bar{P}_{v} = \frac{\pi^2 \mathcal{P}_{v}}{q^3 {R^*}^3 \bar{U}_{f}^2}$ and $z_{\pm} = y(1 \pm c_{\rm s})/2c_{\rm s}$. 
After redshifting, the power spectrum today at frequency $f$ is \cite{Gowling_2021}
\begin{equation}
    \Omega_{\text{GW}} h^2 = 3.57 \times 10^{-5}\left(\frac{100}{g_{*}}\right)^{\frac{1}{3}} \Sigma(v_{\rm w}, \alpha) \mathcal{P}_{\text{GW}}(qR^*(f)),
\end{equation}
where $g_{*}$ is the number of relativistic degrees of freedom, $\Sigma(v_{\rm w}, \alpha)$ is a suppression factor found from recent numerical simulations 
for strong phase transitions~\cite{Cutting:2019zws} and 
\begin{equation}
    f =  2.6 \times 10^{-6}\frac{qR^*}{HR^{*}}\left(\frac{T}{100 \gev}\right)\left(\frac{g_{*}}{100}\right)^{\frac{1}{6}} \mathrm{Hz}.
\end{equation}
In this paper, we set $\Sigma$ as one because it is a constant and does not affect the uncertainty.
\section{Simulation framework and toy quartic model}

To simulate the random generation of bubbles according to Eq.(\ref{eq: Gama}), we follow the method proposed by J. Ignatius \emph{et al.}~\cite{enqvist1992nucleation}

\begin{itemize}
\item[(0)] The phase transition begins when the temperature falls below the critical temperature, so the starting time is set to $t_1=t_{\rm c}$, at the critical temperature $T=T_{\rm c}$. The probability that no bubbles are nucleated in the volume $V_{\rm s}$ during the time interval $t_1 \le t \le t_2$ can be expressed as 
\begin{equation}
    p(t_2, t_1) = {\rm{exp}}\left(-\int_{t_1} ^{t_2} V_{\rm s} \Gamma dt\right).
\end{equation}

\item[(1)] The start time of next bubble nucleated is $t_2$, which is obtained by solving
\begin{equation}
    p(t_2, t_1) = r,
\end{equation}
where $r$ is a uniformly distributed random number between 0 and 1.

\item[(2)] Choose a random location for the bubble within $V_{\rm s}$. If the location falls within a previously generated bubble at $t_2$, reject the bubble and return to step (1).

\item[(3)] Increase the radius of all the bubbles individually by 
\begin{equation}
    R(t) = \int_{t_{\rm birth}}^{t} v_{\rm w}(\tau)d\tau.
\end{equation}

\item[(4)] Replace $t_1$ with $t_2$ and return to step (1) until the bubbles fill the entire space.
\end{itemize}

As is typical in simulations, we adopt the following effective potential
\begin{equation}  \label{eq-29}  
\begin{aligned}
    V(\phi, T) = &\frac{1}{2}M^2(T)\phi^2 - \frac{1}{3}\delta(T)\phi^3 + \frac{1}{4} \lambda \phi^4\\
    =& \frac{1}{2} \gamma (T^2-T_{0}^2) \phi^2 - \frac{1}{3}\alpha T \phi^3 + \frac{1}{4} \lambda \phi^4,
\end{aligned}
\end{equation}
where $T_0$, $\gamma$, $\alpha$, $\lambda$ are adjustable parameters, and here are set to be 75\gev, 0.2, 0.3, 0.2, respectively. This expression can be deduced from the high temperature expansion for Standard Model extensions if no extra scalar field is added.

\begin{figure}[t]
\centering 
\includegraphics[width=0.47\textwidth]{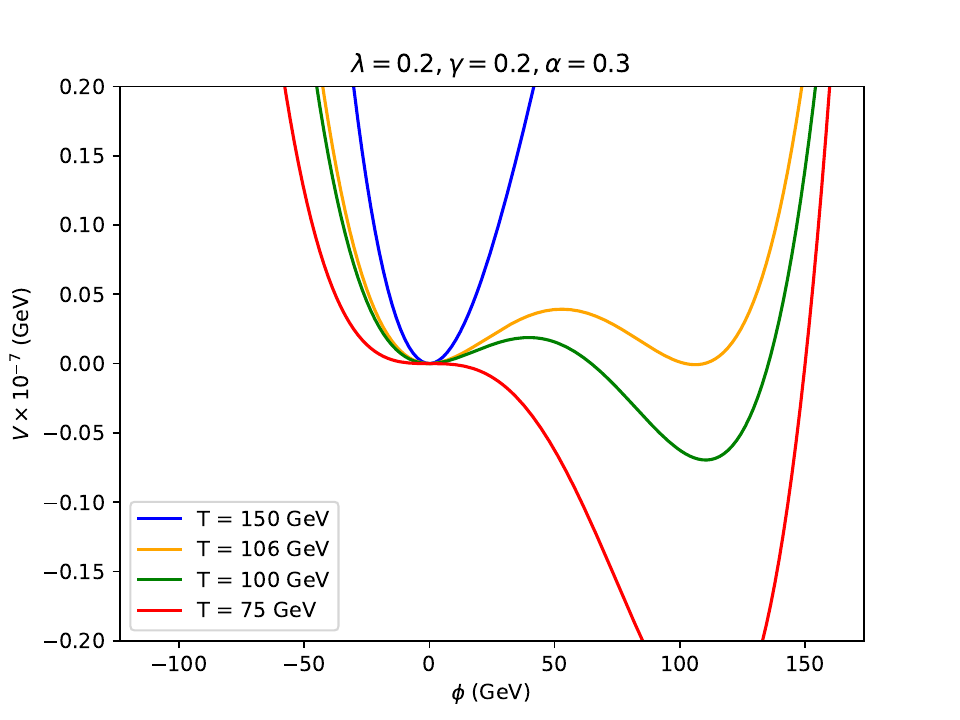}
\vspace{-0.2cm}
\caption{The scaled effective potential for the effective model in Eq.(\ref{eq-29}) varied with temperature when $\lambda$ = 0.2, $\gamma$ = 0.2, $\alpha$ = 0.3. ${T_{\rm c}} \approx 106$ GeV is the critical temperature at which the two minimums have the same potential.  ${T_0} = 75$ GeV is the temperature at which the origin point starts turning from a minimum to a maximum.}
\label{fig:2}
\end{figure}

In the early universe, when the temperature is higher than a threshold $T_0$, i.e. $T \ge T_0$, the potential has a minimum at $\phi = 0$ and $V(\phi=0,T)=0$, indicating that the symmetry is restored. The minimum corresponding to symmetry-broken vacuum appears at low temperature, with an expectation value of
\begin{equation} 
v(T) = \frac{\alpha T}{2\lambda} \left(1 + \sqrt{1-\frac{4 \lambda M(T)^2}{\delta^2}}\right).    
\end{equation}
The critical temperature, at which the two minimums have the same free energy, can be determined as 
\begin{equation}
T_{\rm c} = \sqrt{\frac{T_{0}^2}{1-\frac{2\alpha^2}{9\lambda \gamma}}}.
\end{equation}
For above input parameters, $T_c \approx \sqrt{2} T_0$, which is roughly 106 GeV. When the temperature decreases below $T_0$, the origin becomes a maximum instead of a minimum, and there is no phase transition but rather spinodal decomposition. An example of the variation of the effective potential with temperature is displayed in Fig.~\ref{fig:2}.

% With the effective potential and nucleation rate, we could start the simple bubble expanding simulation by incorporation those ingredients. In the simulation, the bubbles are regarded as simple spheres starting from zero radius and ignoring their internal field configurations. The bubbles expand with the constant velocity (0.3, in this paper) for simplicity. More precise velocity needs to solve the fluid-scalar equation and details could be seen in Ref.~\cite{lewicki2022electroweak,konstandin2014boltzmann,wang2020bubble, megevand2010velocity}. It's worth noting that when the real bubble collision, the true vacuum in the collision region can oscillate and become false vacuum for a while\cite{jinno2019relativistic, gould2021vacuum}. This "trapping" phenomenon has not been considered in Eq~(\ref{eq: h(t)}) and may delay the phase transition completion time. To account for this effect, a real lattice simulation is required, which is very time-consuming and beyond the scope of this paper.

Rather than performing a numerical calculation of the Euclidean action, we utilize the fitted formula proposed by Fred~\cite{adams1993general} (similar fitted formula can be found in Ref.~\cite{Dine:1992wr}),
%To save computing resources, in the simple quartic model we do not directly compute the field configuration and the Euclidean action but adopted a more simple fitting formula obtained by Fred\cite{adams1993general}
\begin{equation} \label{eq: fred fitting}
    S_{\rm E} = \frac{\pi \delta 8\sqrt{2}}{(\lambda/4)^{3/2}243}(2 - e)^{-2}\sqrt{e/2}(\beta_1 e + \beta_2 e^2 + \beta_3 e^3),
\end{equation}
where $\beta_1 = 8.2938$, $\beta_2 = -5.5330$, $\beta_3 = 0.8180$ and $e = 9\lambda M(T)^2/\delta^2$.  
This formula is applicable to a single scalar field $\phi$ within an arbitrary quartic potential and is derived by comparing the real numerical solution and the thin-wall solution. It agrees well with the theoretical asymptotic formulas. It should be noted that the value of $e$ varies between 0 and 2 as the temperature decreases from $T_{\rm c}$ to $T_0$. If the temperature falls below $T_0$, the potential barrier no longer exists and the fitted formula for the Euclidean action will yield incorrect results.

%\begin{figure}[htbp!]
%\centering 
%\includegraphics[width=0.5\textwidth]{action.pdf}
%\caption{Comparison between the Fred's fitting formula Eq~(\ref{eq: fred fitting}) %and the large (small) limit, see Eq (3.3) in Ref.\cite{enqvist1992nucleation}.}
%\label{fig:3}
%\end{figure}

To simulate the bubble expansion, simple spheres are assumed, starting from zero radius, and their internal field configurations are ignored. For simplicity, the bubbles expand at a constant velocity of 0.3 in this paper. A more accurate velocity calculation would require solving the fluid-scalar equation and the details could be seen in Refs.~\cite{lewicki2022electroweak,konstandin2014boltzmann,wang2020bubble, megevand2010velocity}.
It is important to note that during a real bubble collision, the true vacuum in the collision region can oscillate and temporarily become a false vacuum~\cite{jinno2019relativistic,gould2021vacuum}. This trapping" phenomenon has not been taken into account in Eq.(\ref{eq: h(t)}) and may delay the completion time of the phase transition. Accounting for this effect requires a real lattice simulation, which is time-consuming and beyond the scope of this paper.

Following previous lattice simulations, we employ $T_{\rm c}$ as the characteristic quantity and select a cubic simulation volume with its length $L_{\rm s}$ in the range of $0.2\% /H(T_{\rm c}) \sim 0.8\% /H(T_{\rm c})^3$, where $1/H(T_{\rm c})$ represents the Hubble size at $T_{\rm c}$. When a bubble expands into the boundary, the part of the bubble that passes the boundary is ignored.  With those volumes, approximately 500 $\sim$ 22000 bubbles can be generated using the aforementioned method. In practice, the phase transition occurs over an exceedingly brief duration of approximately $10^{-11}$s, which means that the interval between bubble generation is extremely short. To avoid  numerical errors, we substitute time with temperature by 
\begin{equation} \label{eq: dT/dt}
    \frac{dT}{dt} = -H(T)T,
\end{equation}
where $H$ is the Hubble constant. We assume that the phase transition takes place  during the radiation dominant period, thus implying
\begin{equation}
    H(T) = \sqrt{\frac{8 \pi G \rho_r(T)}{3}} = \sqrt{\frac{8 \pi^3 G g T^4}{90}} ,
\end{equation}
where $g = 106.75$ is the number of relativistic degrees of freedom and $G = 6.71 \times 10^{-39} \gev^{-2}$ is the Newton’s gravitational constant. With this equation, $\beta_{\rm th}$ can be correlated with temperature. When the pre-factor is constant $T_{\rm c}^4$, $\beta_{\rm th}$ can be written as
\begin{equation}
    \beta_{\rm th} = -\frac{{\rm d}{\rm ln} T_{\rm c}^4}{{\rm d}t} + \left. \frac{{\rm d}(S_{E}/T)}{{\rm d} t}\right|_{t=t_f} = \left. \frac{{\rm d}(S_{E}/T)}{{\rm d} T} \frac{{\rm d} T}{{\rm d} t}\right|_{t=t_f},
\end{equation}
where $t_f$ can be calculated by using $h_{\rm th}(t_f) = 1/e$. 
Based on the parameters we have selected, its value is about $4.59 \times 10 ^{-11} \gev$ and $\beta_{\rm th}/H(T_{\rm c}) \approx 2908$. We also utilize $\beta_{\rm th}^{-1}$ to measure the length of the cubic simulation volume, so that our $L_{\rm s}$ are about 5.82$\beta_{\rm th}^{-1}$, 8.73$\beta_{\rm th}^{-1}$, 14.54$\beta_{\rm th}^{-1}$ and 23.27$\beta_{\rm th}^{-1}$ for $0.2\% /H(T_{\rm c})$, $0.3\% /H(T_{\rm c})$, $0.5\% /H(T_{\rm c})$ and $0.8\% /H(T_{\rm c})$, respectively. Furthermore, $\Gamma_0$ is about 1.7 $\beta_{\rm th}^4$. 

We utilize the Monte Carlo method to accurately determine the fraction of false vacuum. More precisely, we divide the temperature into 3000 intervals and, in each interval, we randomly distribute 15,000 points throughout the volume. We then count the number of points that lie outside the bubbles, which allows us to precisely track the variation of the false vacuum fraction with time.

\section{Results and Discussions}

\subsection{Nucleation Rate}

We firstly verify the validity of the simulation procedure outlined above by checking if the nucleation rate agrees with Eq.(\ref{eq: Gama}). 
By dividing the temperature range for bubble generation into several intervals, the nucleation rate per unit time per unit volume in the simulation can be determined by
%After generating enough numbers of bubbles, we could verify if the simulation is successful. Dividing the bubbles generation temperature into several parts and denoting each interval length and the corresponding bubble number generated in this interval is $\Delta T$ and $\Delta N$, then the nucleation rate per time per volume could be written as 
\begin{equation} \label{eq: simulation gama}
    \Gamma_{\rm sim}=\frac{\Delta N}{V_{\rm false}\Delta t} = \frac{\Delta N}{V_{\rm false}\Delta T} \frac{\Delta T}{\Delta t} \approx \frac{\Delta N}{V_{\rm false}\Delta T} \frac{dT}{dt},
\end{equation}
where $\Delta T$ denotes the length of each interval and $\Delta N$ is the corresponding number of bubbles generated in that interval. Then, it can be compared to the analytical formula Eq.(\ref{eq: Gama}).

\begin{figure}[htbp!]
\centering 
\includegraphics[width=0.47\textwidth]{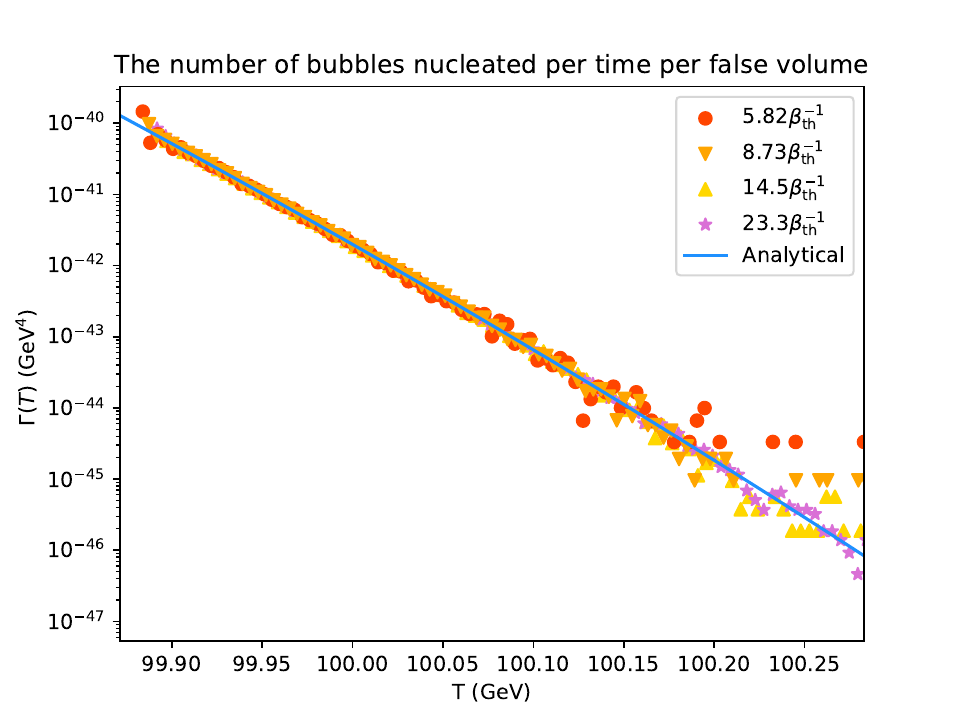}
\vspace{-0.2cm}
\caption{Comparison between $\Gamma$ calculated with Eq.(\ref{eq: Gama}) and $\Gamma_{\rm sim}$ measured from simulation. The red, gold, yellow and purple points are the simulation data with $L_{\rm s} = $ 5.82$\beta_{\rm th}^{-1}$, 8.73$\beta_{\rm th}^{-1}$, 14.54$\beta_{\rm th}^{-1}$ and 23.3$\beta_{\rm th}^{-1}$. The blue line is the analytical formula Eq.(\ref{eq: Gama}) for pre-factor1. For each scatter point, the corresponding $T$ value is the center value of the interval.}
\label{fig:4}
\end{figure}

In Fig.~\ref{fig:4}, We presented the average results of $\Gamma_{\rm sim}$ of 50 simulations under different volumes for pre-factor 1 and compared them with the theoretical formula Eq.(\ref{eq: Gama}). The comparisons show that  Eq.(\ref{eq: Gama}) and Eq.(\ref{eq: simulation gama}) generally agree well in most temperature intervals, though deviations are observed at both high and low temperatures. During the early stage of the phase transition, only a few bubbles are randomly generated, which leads to strong statistical fluctuations and in turn results in the deviations at high temperatures. When $T \lesssim 99.88~\gev$, the corresponding value of $\Gamma_{\rm sim}$ diverges and cannot be displayed in the figure. This is because, in the final stage of the phase transition, almost all the region is filled with bubbles. As a result, the theoretical false vacuum fraction approaches zero, but the Monte Carlo simulations give an exact zero, leading to a divergence in Eq.(\ref{eq: simulation gama}). 

By comparing the distribution of four simulations, we observe that a larger simulated volume could mitigate statistical fluctuations during the initial phase of the transition. Additionally, we can accumulate more simulations to further diminish these fluctuations.
However, due to the consumption of computing resources and the fact that $\Gamma_{\rm sim}$ is in agreement with $\Gamma$ most of the time, it is safe to use above Monte Carlo method to study the statistical measures that characterize bubble kinematics. 
Thus, except for such special regions, the nucleation rate measured from simulations indeed agree with Eq.(\ref{eq: Gama}).

\begin{table}[]
\caption{Results of the total number of bubbles of pre-factor 1 from simulation and analytical formula}
\label{tab:1}
\begin{tabular}{lcccc}
\hline
\diagbox{$N_{\rm tot}$}{$L_{\rm s}$}  & 5.82$\beta_{\rm th}^{-1}$ & \multicolumn{1}{l}{8.73$\beta_{\rm th}^{-1}$} & \multicolumn{1}{l}{14.5$\beta_{\rm th}^{-1}$} & \multicolumn{1}{l}{23.3$\beta_{\rm th}^{-1}$} \\
\hline
Max $N_{\rm tot}^{\rm sim}$ & 573              & 1627                                  & 6373                 &    23480        \\
Min $N_{\rm tot}^{\rm sim}$ & 282              & 1113                                  & 5222                &      21341         \\
Avg $N_{\rm tot}^{\rm sim}$ & 426              & 1337                                  & 5709                  &     22378       \\
$N_{\rm tot}^{\rm th}$            & 324              & 1092                                  & 5059               &  20724             \\
$N_{\rm tot}^{\rm hybrid}$                  & 427              & 1347                                  & 5722      & 22380  \\       
$N_{\rm tot}^{\rm approx}$                  & 290              & 979                                  & 4533      & 18567  \\ 
\hline
\end{tabular}
\end{table}

\subsection{Total number of bubbles}
The first statistical measure we study is the total number of bubbles that are generated during the phase transition.
Table~\ref{tab:1} lists the total number of bubbles of pre-factor1 obtained from the analytical formula and from simulation with four different simulated volume.
The first three rows show the maximum, minimum, and average numbers of bubbles generated from 50 simulations, respectively.
The fourth row shows the theoretical total number of bubbles results obtained from Eq.(\ref{eq: total number})

\begin{equation}
    N^{\rm th}_{\rm tot} =  \int_{t_{\rm c}} ^{t_0} V_{\rm s} \Gamma(t') h_{\rm th}(t') dt',
\end{equation}
where $t_0$ is the time corresponding to $T_0$. The fifth row shows the same results as the fourth row, except that the false vacuum fraction $h$ in Eq.(\ref{eq: total number}) is replaced by $h_{\rm sim}$. Its expression is
\begin{equation}
    N^{\rm hybrid}_{\rm tot} =  \int_{t_{\rm c}} ^{t_0} V_{\rm s} \Gamma(t') h_{\rm sim}(t') dt'.
\end{equation}
The bottom row displays the results of popular formula
\begin{equation} \label{eq: N_approx}
    N^{\rm approx}_{\rm tot} = \frac{\beta_{\rm th}^3 V_{\rm s}}{8 \pi v^3}, 
\end{equation}
where the approximate nucleation rate Eq.(\ref{eq: approx nucleation rate}) is used.

It can be seen that the numbers of bubbles generated in different simulated volumes vary significantly. By comparing the Avg $N_{\rm tot}^{\rm sim}$ with the $N_{\rm tot}^{\rm th}$, it is apparent that they do not align closely. 
The discrepancies, amounting to 23.9\%, 18.3\%, 11.3\% and 7.39\% for increasing simulated volumes, align with our intuition that as the simulated volume increases, the corresponding results tend to converge towards the theoretical predictions.
Besides, we observe that differences between $N_{\rm tot}^{\rm approx}$ and $N_{\rm tot}^{\rm th}$ are significantly larger. 
Specifically, the errors are 31.9\% and 20.6\% for $L_{\rm s} =  5.82\beta_{\rm th}^{-1}$ and $L_{\rm s} = 14.5\beta_{\rm th}^{-1}$, respectively. 
In these simulated volumes, the total number of bubbles almost comparable to those in previous lattice simulations~\cite{hindmarsh2015numerical, hindmarsh2014gravitational,hijazi2022numerical}, raising the question of to what extent will lattice results be affected if the approximate formula Eq.(\ref{eq: N_approx}) is used in simulating the nucleation process of bubbles. 
In Ref.\cite{hijazi2022numerical}, $N(t)/N_{\rm tot}$ is regarded as a good approximation for true vacuum fraction. By using Eq.(\ref{eq: simpifiled h}), we obtain
\begin{align}
    \frac{N}{N_{\rm tot}^{\rm approx}} &= 1 - {\rm exp}\left[-e^{\beta(t-t_f)}\right] \notag, \\
    t_n &= \frac{1}{\beta}{\rm ln}(-{\rm ln}(1- \frac{N}{N_{\rm tot}^{\rm approx}})) + t_f,
\end{align}
where $t_n$ is the nucleation time of the n-th bubble. Based on our findings, it is important to note that the nucleation process described above should be used cautiously when the simulated volume is small, despite being a convenient estimation of the nucleation time. For the other two pre-factors, simulations were conducted with a simulated length of 14.5 $\beta_{\rm th}^{-1}$, revealing that the differences between $N_{\rm tot}^{\rm th}$ and $N_{\rm tot}^{\rm sim}$ are also around 10\%. Specifically, $N_{\rm tot}^{\rm sim}$ for pre-factor 3 is 6375, exceeding that of pre-factor 1 by roughly 10\%. However, $N_{\rm tot}^{\rm sim}$ for pre-factor 2 is 5709, remains comparable to pre-factor 1.
%%fig.4 
\begin{figure*}[htbp!]
\centering 
\includegraphics[width=1.0\textwidth]{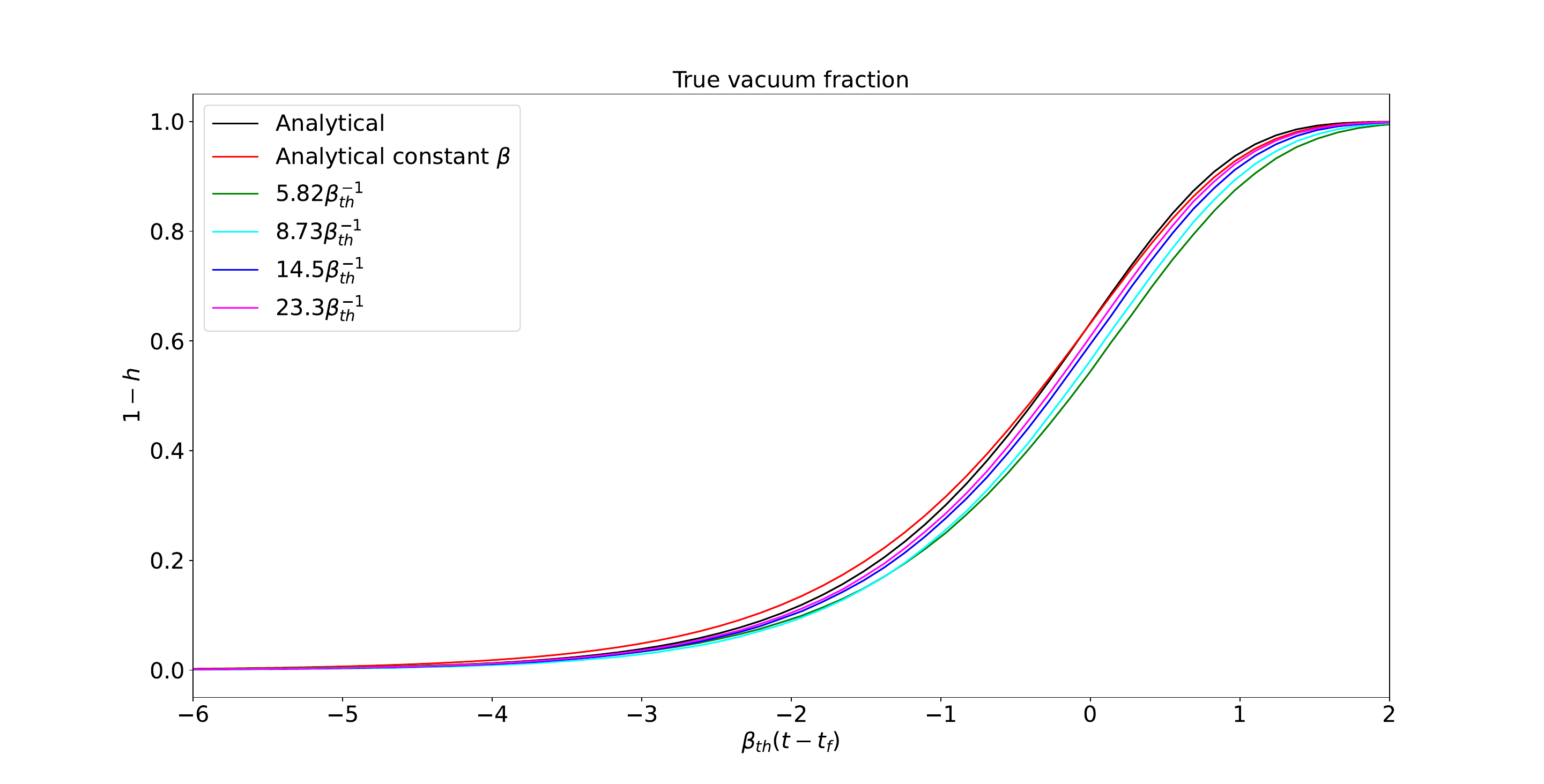}
\vspace{-0.7cm}
\caption{Comparison between the analytical $h$ and $h_{\text{sim}}$ of varying simulated volumes. The green, cyan, blue and purple curves are obtained by interpolating the stacked simulation data with increasing simulated volume. The black curve represents the outcome of Eq.(\ref{eq: h(t)}), while the red curve depicts the result of Eq.(\ref{eq: simpifiled h}).}
\label{fig:5}
\end{figure*}

For all the simulated volumes, the average simulation results $N_{\rm tot}^{\rm sim}$ show good agreement with $N_{\rm tot}^{\rm hybrid}$. The only difference between the calculation of $N_{\rm tot}^{\rm th}$ and $N_{\rm tot}^{\rm hybrid}$ is the false vacuum fraction employed. Following Ref.\cite{enqvist1992nucleation}, we present the true vacuum fraction of both the analytic and simulation results for pre-factor 1 in Fig.~\ref{fig:5}, using the dimensionless scaled time $\beta_{\rm th}(t-t_f)$ as the x-axis to better illustrate the differences under varying conditions.

It is observed that as the simulated volume increases, the simulated results tend to converge towards the theoretical results. To quantitatively capture this trend, we utilize the theoretical result $h_{\rm th}$ as a reference and calculate the averaged absolute and relative errors of the simulation results,
\begin{align}
    \sigma_{\rm abs}^{i} &= \sum_{j}^{L}\frac{\left|h_{\rm sim}^{i}(t_j) - h_{\rm th}(t_j)\right|}{L}, \notag\\
    \sigma_{\rm rel}^{i} &= \sum_{j}^{L}\frac{\left|h_{\rm sim}^{i}(t_j) - h_{\rm th}(t_j)\right|}{h_{\rm th}(t_{j})L},
\end{align}
where $i$ stands for different simulated length results, $L$ is the number of the averaged data. 
The absolute errors are 0.0550\%, 0.0473\%, 0.0259\%, 0.0157\% for the increasing simulated volume, with corresponding relative ones of 9.28\%, 8.98\%, 7.98\%, 8.13\%. 
The absolute error $\sigma_{\rm abs}$ between $h_{\rm th}$ and $h_{\rm approx}$ is about 0.0155\%, primarily due to the Taylor expansion and deviations from the assumptions regarding the Euclidean action. However, the relative error $\sigma_{\rm rel}$ between $h_{\rm th}$ and $h_{\rm approx}$ is approximately 5753\%, mainly because the $h_{\rm approx}$ can not precisely capture the behaviour of the $h_{\rm th}$ in the early stage of phase transition. During this time, The $h_{\rm approx}$ is on the order of  $\mathcal{O}(10^{-13})$ while the $h_{\rm th}$ is about $\mathcal{O}(10^{-16})$. Although we can not distinguish them in the Fig.\ref{fig:5}, the relative error remains significantly large . If we disregard the early stage by imposing the condition $h > 10^{-5}$, the relative error then decreases to 13\%, yet it remains higher than that of the simulated results.  This highlights the importance of exercising caution when utilizing the simplified formula.

For subsequent calculations of the gravitational wave spectrum, we need to select a reference temperature. Here we set the reference temperature $T_f$ as the time near the final stage of the phase transition, specifically when the false vacuum fraction reaches $1/e$. Since $h_{\rm th}$ and $h_{\rm approx}$ are indistinguishable in this interval, the corresponding $T_f$ values are virtually identical, about 99.959 \gev. The simulated $T_f$ are 99.952 \gev, 99.954 \gev, 99.956 \gev and 99.958 \gev for increasing simulated volume. This indicates that the bubble kinetics at this critical time are accurately captured for all simulated volumes.

This similarity holds true for the other two pre-factor modes with $L_{\rm s } = 14.5\beta_{\rm th}^{-1}$. The simulation-derived reference temperatures for pre-factor2 and 3 are respectively 99.951~\gev and 100.073~\gev, and the analytic reference temperatures are respectively 99.953~\gev and 100.078~\gev.   

%%fig.5
\begin{figure}[htbp!]
\centering 
\includegraphics[width=0.48\textwidth]{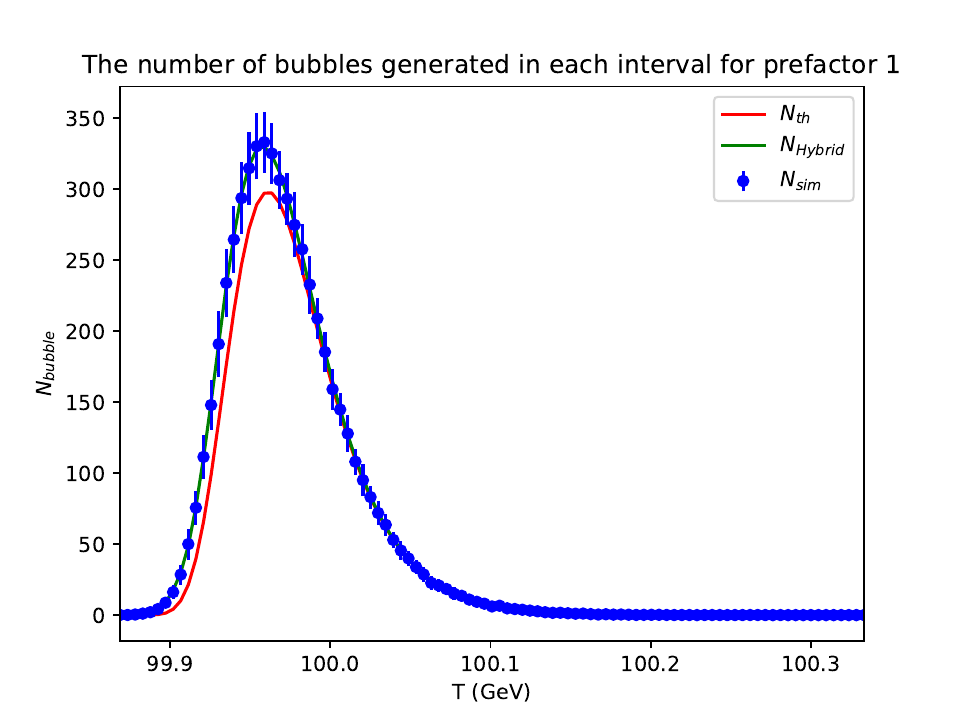}
\vspace{-0.7cm}
\caption{The distribution of bubble numbers generated in each interval $\left[T_i, T_{i+1}\right]$ ($T_{i+1} \ge T_i$) . The blue points with error bar are from simulations, 
the red curve corresponds to the analytical formula $N_{\rm tot}^{\rm th}(T_i) -N_{\rm tot}^{\rm th}(T_{i+1})$ and the green one is obtained similarly as the blue one but with the false vacuum fraction replaced by the measured value from simulations $h_{\text{sim}}$.}
\label{fig:6}
\end{figure}

With these results, we can roughly see the reasons underlying the differences between $N_{\rm tot}^{\rm th}$ and $N_{\rm tot}^{\rm sim}$. Firstly, it assumes an infinite volume in he derivation of $h_{\rm th}$. However, in calculating $N_{\rm tot}^{\rm th}$, we must employ a finite volume $V_{\rm s}$ to obtain a finite result. This inconsistency with the initial assumption introduces an error. It is similar to the thermodynamic limit, where the number density remains finite, but both the number itself and the volume are infinite.

To gain a deeper understanding of why the artificial error in the false vacuum fraction can cause significant differences in the total number of bubbles, we take the results with $L_{\rm s} =  14.5 \beta_{\rm th}^{-1}$ as example and illustrate the distribution of the number of bubbles generated within each temperature interval in Fig.~\ref{fig:6}. The blue point with error bar indicates the expected number $N_{\rm tot}^{\rm sim}(T_i) -N_{\rm tot}^{\rm sim}(T_{i+1})$ from simulation, while the red and green curves are calculated by analytical and hybrid approaches, respectively. 
We can see that the green curve aligns more closely with the  simulation results than with the results obtained from the analytical formula.
At temperatures above 100 GeV (about -1.33 in terms of $\beta_{\rm th}(t-t_f)$), the simulation results and theoretical predictions exhibit good agreement. However, as the temperature decreases, deviations begin to emerge, reaching a peak at approximately 99.96 GeV. This deviations persist until no bubbles are produced at all. Specifically, in the late stage of the phase transition, this error has exceeded two standard deviation, indicating our results have a certain level of statistical significance.

Combing the result of Fig.~\ref{fig:5} and Fig.~\ref{fig:6}, we find that the disparity in the total number of bubbles is attributed to the slight variations in $h$ and $\Gamma$, which are magnified by $dt/dT$. As shown in Fig.~\ref{fig:5}, the discrepancy between $h_{\rm th}$ and $h_{\rm sim}$ becomes apparent as the temperature decreases to 100.2 GeV. However, this deviation is heavily suppressed because of large action, resulting in no disparity in the number of bubbles generated during the initial stage. As the temperature continues to decrease slightly, the action undergoes rapid changes, so the difference between $h_{\rm sim}$ and $h_{\rm th}$ increases, as well as the number of bubbles generated. The difference between $h_{\rm th}$ and $h_{\rm sim}$ reaches its peak at about 99.96 GeV, resulting in a significant disparity in the number of bubbles. Accordingly, we can also infer that the difference in the total number of bubbles may increase if the phase transition occurs in a temperature region where the action is relatively small and changes gradually.
 
\begin{figure}[htbp!]
\centering 
\includegraphics[width=0.48\textwidth]{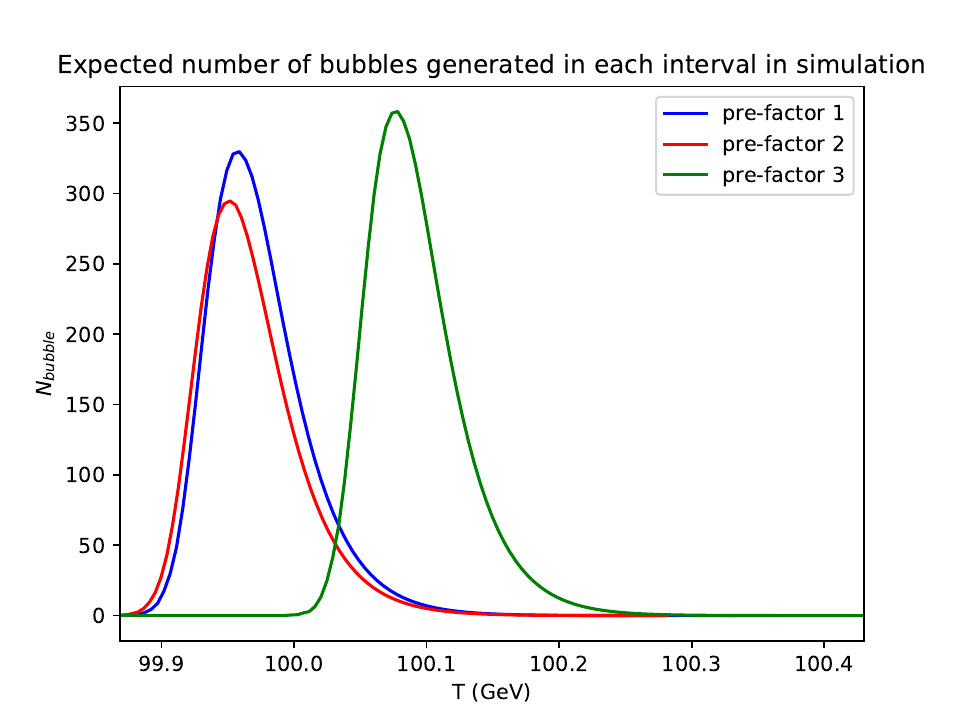}
\vspace{-0.7cm}
\caption{Comparison between the expected numbers of bubbles generated in each interval in simulation for three pre-factors. The blue, red and green curves are for pre-factor1, 2 and 3, respectively.}
\label{fig: three pre-factors}
\end{figure}

As the $N^{\rm hybrid}$ agrees with $N^{\rm sim}$ very well in any small interval, we could regard the $N^{\rm hybrid}$ as the $N^{\rm sim}$ and extract them to assess the impact of different pre-factors with $L_s  = 14.5 \beta_{\rm th}^{-1}$ in Fig.~\ref{fig: three pre-factors}. The blue, red and green curves represent the expected numbers of bubbles generated in each interval for pre-factor1, 2, and 3, respectively. Compared to the pre-factor1 and 2, the phase transition occurred approximately 0.12 GeV earlier for pre-factor3. The maximum value of bubble number for pre-factor3 is approximately 60 larger than pre-factor2, and about 20 larger than pre-factor1.
%If the reference temperature of gravitational waves is the temperature at which the phase transition is completed, pre-factor mode-3 will provide more insights into the earlier universe \ghk{What does this mean?}. 

%%%fig.6 

\begin{figure}[htbp!]
\centering 
\includegraphics[width=0.48\textwidth]{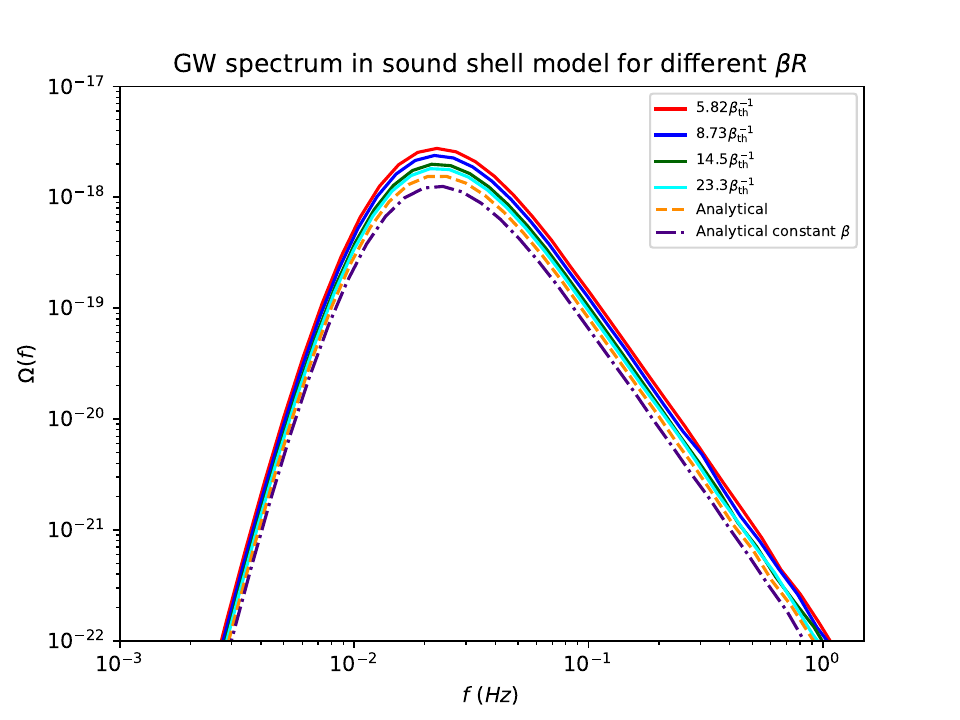}
\vspace{-0.7cm}
\caption{Gravitational wave spectra for different $\beta R^{*}$. The dashed dotted curves are the results of the original sound shell model, the dashed curves are the results with $\beta_{\rm th}R^{*}_{\rm th}$ and the solid curves are the results with $\beta_{\rm sim} R^{*}_{\rm sim}$. The results of $\beta_{\rm th}R^{*}_{\rm th}$ and original sound shell model are almost independent with the simulated volume.}
\label{fig: different betaR}
\end{figure}

As the small difference of false vacuum fraction will result in a non-negligible difference in the total number of bubbles and consequently affect 
the bubble mean separation $R^{*}$, we expect it could accordingly change the gravitational wave spectra. The calculations of the gravitational wave spectra are done in the sound shell model for pre-fractor1, and the results are presented in Fig.~\ref{fig: different betaR}. There the dashed curves denote the spectra obtained using $\beta_{\rm th}R^{*}_{\rm th}$ and $\beta_{\rm approx} R^{*}_{\rm approx}$ while the solid curves are the results of using $\beta_{\rm sim} R^{*}_{\rm sim}$. It is observed that there is an enhancement in gravitational wave spectra. Specifically, as the simulated volume increases, the ratio of spectral peak values with simulated parameter to the theoretical (theoretical approximate) one increased by 1.79 (2.21), 1.54 (1.90), 1.29 (1.59) and 1.17 (1.45) times. All the theoretical results are almost independent with the simulated volume.

We also check the peak value of gravitational wave spectra for other pre-factors with simulated length of 14.5 $\beta_{\rm th}^{-1}$ and find that it increase by a factor of 1.28 and 1.25 compared with the theoretical results. 

%%%fig.7 

\subsection{Bubble Lifetime Distribution}
The gravitational wave spectra can also be modified by the change in the bubble lifetime distribution, which can be measured directly from
simulations and compared with the corresponding analytical formula.
To measure it from simulations, we compute the destruction temperature $T_{\rm death}$ for each bubble (the temperature at which the nucleation site of the bubble is first occupied by the phase boundary formed by other bubbles.), and define the lifetime for a bubble as 
\begin{equation}
    \bar{t} = \hat{t}(T_{\rm death}) - \hat{t}(T_{\rm birth}),
\end{equation}
where $\hat{t}$ represents the solution of the differential equation described in Eq.(\ref{eq: dT/dt}).
By dividing the interval $\left[\min(\beta \bar{t}), \max(\beta \bar{t})\right]$ into many small intervals $\left[\beta \bar{t}_{i}, \beta \bar{t}_{i} + \beta \Delta \bar{t}\right]$, where 
\begin{equation}
    \beta = \left. -\frac{d}{dt} \ln\Gamma\right|_{t=\hat{t}(T_f)},
\end{equation}
we can count the number of bubbles $N_i$ in each interval. Thus the probability density for a bubble having the lifetime $\bar{t}_i$ is
\begin{equation}
    \nu_i = N_i/(N_{\rm tot} \beta \Delta \bar{t}).
\end{equation}

%%fig.8  
\begin{figure}[htbp!]
\centering 
\includegraphics[width=0.48\textwidth]{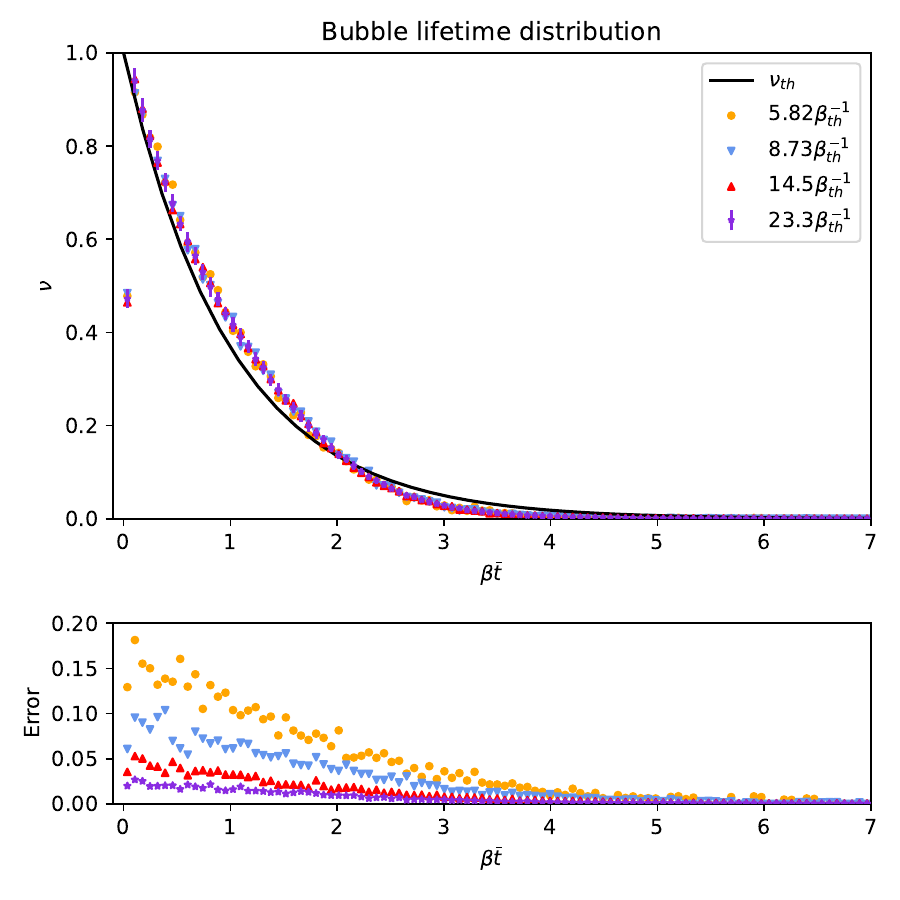}
\vspace{-0.7cm}
\caption{The distribution of bubble lifetime. In upper panel, the black line is the analytical formula Eq.(\ref{eq: lifetime}) and the scatter points are the simulation results, in which the orange one is for simulated length $L_{\rm s} = 5.82\beta_{\rm th}^{-1}$, the blue one is for simulated length $L_{\rm s} = 8.73\beta_{\rm th}^{-1}$, the red one is for simulated length $L_{\rm s} = 14.5\beta_{\rm th}^{-1}$ and the violet one is for simulated length $L_{\rm s} = 23.3\beta_{\rm th}^{-1}$. In lower panel, the errors are displayed. The $\beta \bar{t}$ for each rectangle is the center value of the interval.}
\label{fig:7}
\end{figure}

Fig.~\ref{fig:7} shows the bubble lifetime distribution for pre-factor1 with different simulated volume.
The black curve represents Eq.(\ref{eq: lifetime}), while the scatter points represents the simulation result. We see that the error of each simulated data is rapidly decreasing with increasing of simulated volume and the theoretical results do not exhibit good agreement with the simulated ones. Eq.(\ref{eq: lifetime}) tends to underestimate the distribution of bubble lifetimes when $0.14 \le \beta \bar{t} \le 2$, while overestimating the distribution in other regions. This suggests that the bubble lifetime distribution is not an ideal exponential distribution as described by Eq.(\ref{eq: lifetime}). This deviation could be explained by noting that, in the derivation of $\nu_{\rm th}$, a constant distance $L$ is utilized. However, it is actually possible for a bubble to spontaneously appear between the phase boundary and the bubble we are studying, leading to a sudden decrease in $L$ and resulting in a shorter lifetime. As a result, the number of long-lived bubbles is suppressed, and their corresponding count is transferred to the range of shorter-lived bubbles ($0.14 \le \beta \bar{t} \le 2$). Furthermore, due to the normalization of $\nu$, the area below $\nu$ equals to 1.  Thus, this transferred counting will result in a decrease in the areas of the near zero interval. Similar findings are also reported in Ref.\cite{hijazi2022numerical}, where the definition of bubble lifetime slightly differs. To make sure this is not due to a numerical artifact, we examined the bubble birth time within near zero interval and observed that it is not concentrated within a particular range.

\begin{figure}[htbp!]
\centering 
\includegraphics[width=0.48\textwidth]{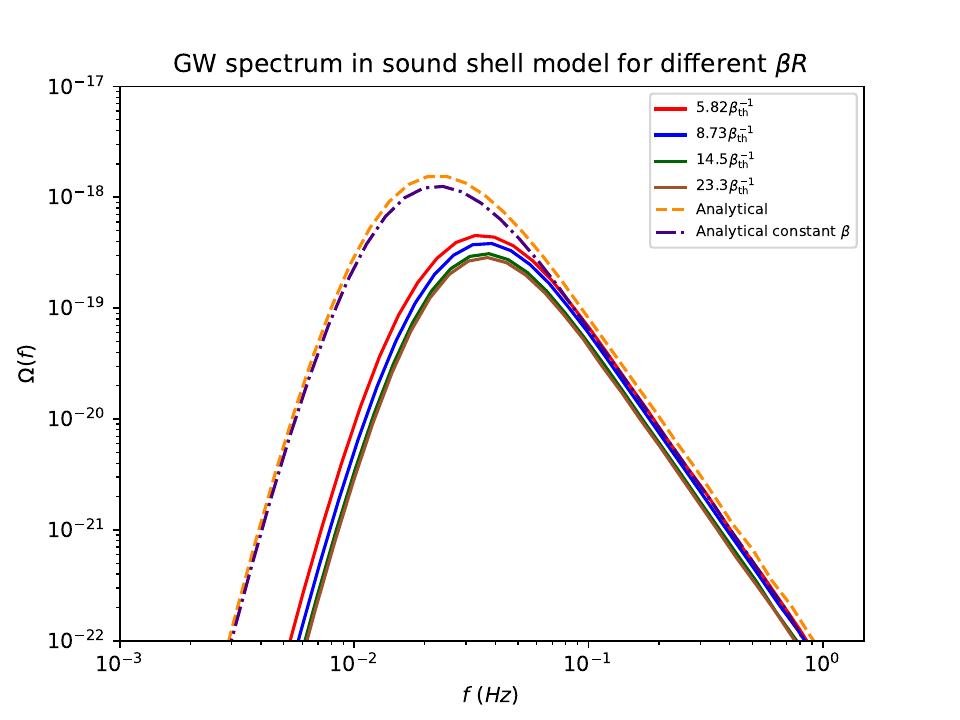}
\vspace{-0.7cm}
\caption{The gravitational wave spectra calculated by combining our simulation results and sound shell model. The dashed curves are the results of the theoretical spectra and the solid curves are obtained with simulation results. Here we set the alpha factor in front of the bubble wall $\alpha_{+}$ as 0.0046~\cite{hindmarsh2019gravitational}.}
\label{fig:8}
\end{figure}

Given the distinctive behavior of the bubble lifetime distribution, we anticipate that it will result in a modification of gravitational wave spectra. By inserting our simulation results $R^{*}_{\rm sim}$ and $\nu_{\rm sim}$, we get the final gravitational wave spectra and compare it with the original spectra in Fig.~\ref{fig:8}. The dashed curves in this figure are obtained with the original set of analytical inputs while the solid curves are obtained with inputs from numerical simulations where the bubble lifetime distribution is replaced by $\nu_{\rm sim}$. 

%%%fig.9 

From Fig.~\ref{fig:8} we find that: (1) the gravitational wave spectra are almost the same for three different pre-factors; (2) the peak of the spectra shifts to a higher frequency compared with the original spectra and the original one overestimates the gravitational wave in the whole range of $f$. This suppression effect could be explained by noting that $\nu$ and $|A|^2$ are all rapidly decreasing to a small number when the independent variable deviated from the origin~\cite{guo2021phase} and so the main contribution to $\mathcal{P}_{v}$ comes from a small $f$ (which is equivalent to the $qR^{*}$) and small $\beta \bar{t}$, which exactly corresponds to one of the suppression interval in Fig.~\ref{fig:7}.

\begin{table}[]
\caption{Deviation of characteristic physical quantities}
\begin{tabular}{lllll}
\hline
Uncertainty & $5.82\beta_{\rm th}^{-1}$ & $8.73\beta_{\rm th}^{-1}$ & $14.54\beta_{\rm th}^{-1}$ & $23.3\beta_{\rm th}^{-1}$ \\
\hline
$T_{f}$       & 0.007\%     & 0.005\%     & 0.003\%    & 0.001\% \\
$\beta R^{*}$         & 13.9\%      & 11.1\%      & 7.99\%     &6.42\%  \\
$N_{\rm tot}$       & 31.9\%     & 26.8\%     & 20.6\%     & 17.0\%\\
% $f^{\rm peak}_{\beta R^{*}}$  & 6.15\%           & 8.77\%            & 11.7\%         &* &* \\ 
$\Omega_{\rm GW}h^2_{\beta R^{*}}$  & 54.6\%           & 47.4\%           & 36.9\%         & 31.1\%\\ 
% $f^{\rm peak}_{\rm sim}$  & 6.15\%           & 8.77\%            & 11.7\%         & *&* \\ 
$\Omega_{\rm GW}h^2_{\nu}$  & 177\%           & 228\%           & 304\%        & 338\%  \\ 
\hline
\end{tabular}
\end{table}

Finally, we list the deviations of some characteristic physical quantities $|(C_{\rm approx} - C_{\rm sim})| / C_{\rm sim}$ in Table II, where the row $\Omega_{\rm GW}h^2_{\beta R^{*}}$ shows the corresponding uncertainty of the gravitational wave spectra peak values with different $\beta R^{*}$. The rows and $\Omega_{\rm GW}h^2_{\nu}$ are similar with the previous one.

\section{Conclusion}

In this work we conduct simulations of bubble nucleation and compare the results with the usually adopted analytical formula. Our major findings are the following:
\begin{enumerate}
    \item As the simulated volume increases, the simulated false vacuum fraction  approaches to the theoretical one. The simulated results rarely change when the simulated length $L_{\rm s}$ exceeds $14.5 \beta_{\rm th}^{-1}$.
    
    \item Since the theoretical formula for the false vacuum fraction presupposes an infinite volume, the total number of bubbles calculated using this formula for a finite volume diverges from the simulated value. However, this discrepancy diminishes as the simulated volume expands.
    
    \item Different pre-factors do not lead to significantly differences in bubble kinetics.
    
    \item The actual bubble lifetime distribution is similar with the previous results of Ref.\cite{hijazi2022numerical} and does not follow an exponential distribution, which results in a suppression effect on the gravitational wave spectra across the entire frequency range.
\end{enumerate}

We hope our work can contribute to the precise detection of the spectrum of gravitational waves in the future. Additionally, there are many uncertainties that need to be measured, such as errors caused by the non-uniform expansion of bubbles. We will systematically estimate these uncertainties in our subsequent works.

\section*{Acknowledgments}
Y.X thanks De Yu Wang for helpful discussion. This work was supported by the National Natural Science Foundation of China (12105248, 11821505, 12075300, 12335005, 12147103),  by Peng-Huan-Wu Theoretical Physics Innovation Center (12047503),
and by the Key Research Program of the Chinese Academy of Sciences (XDPB15).

% \section*{Appendix}
% In Figs.~\ref{fig:10}-\ref{fig:13} we compare the simulated results with the theoretical values on the bubble nucleation rate, the false vacuum fraction, the bubble number generated in small intervals and the bubble lifetime distribution for pre-factor 2 and 3.

% \begin{figure}[!t]
% \centering 
% \includegraphics[width=0.4\textwidth]{bubble_lifetime2.pdf}\\
% \includegraphics[width=0.4\textwidth]{bubble_lifetime3.pdf}
% \caption{Same as Fig.~\ref{fig:7}, but for pre-factor2 and pre-factor3.}
% \label{fig:13}
% \end{figure}

 %\clearpage
 
\bibliography{apssamp}% Produces the bibliography via BibTeX.

\providecommand{\noopsort}[1]{}\providecommand{\singleletter}[1]{#1}%
\begin{thebibliography}{66}
\expandafter\ifx\csname natexlab\endcsname\relax\def\natexlab#1{#1}\fi
\expandafter\ifx\csname bibnamefont\endcsname\relax
  \def\bibnamefont#1{#1}\fi
\expandafter\ifx\csname bibfnamefont\endcsname\relax
  \def\bibfnamefont#1{#1}\fi
\expandafter\ifx\csname citenamefont\endcsname\relax
  \def\citenamefont#1{#1}\fi
\expandafter\ifx\csname url\endcsname\relax
  \def\url#1{\texttt{#1}}\fi
\expandafter\ifx\csname urlprefix\endcsname\relax\def\urlprefix{URL }\fi
\providecommand{\bibinfo}[2]{#2}
\providecommand{\eprint}[2][]{\url{#2}}

\bibitem[{\citenamefont{Abbott et~al.}(2016)\citenamefont{Abbott, Abbott, Abbott, Abernathy, Acernese, Ackley, Adams, Adams, Addesso, Adhikari et~al.}}]{abbott2016observation}
\bibinfo{author}{\bibfnamefont{B.~P.} \bibnamefont{Abbott}}, \bibinfo{author}{\bibfnamefont{R.}~\bibnamefont{Abbott}}, \bibinfo{author}{\bibfnamefont{T.}~\bibnamefont{Abbott}}, \bibinfo{author}{\bibfnamefont{M.}~\bibnamefont{Abernathy}}, \bibinfo{author}{\bibfnamefont{F.}~\bibnamefont{Acernese}}, \bibinfo{author}{\bibfnamefont{K.}~\bibnamefont{Ackley}}, \bibinfo{author}{\bibfnamefont{C.}~\bibnamefont{Adams}}, \bibinfo{author}{\bibfnamefont{T.}~\bibnamefont{Adams}}, \bibinfo{author}{\bibfnamefont{P.}~\bibnamefont{Addesso}}, \bibinfo{author}{\bibfnamefont{R.}~\bibnamefont{Adhikari}}, \bibnamefont{et~al.}, \bibinfo{journal}{Physical review letters} \textbf{\bibinfo{volume}{116}}, \bibinfo{pages}{061102} (\bibinfo{year}{2016}).

\bibitem[{\citenamefont{Weir}(2018)}]{weir2018gravitational}
\bibinfo{author}{\bibfnamefont{D.~J.} \bibnamefont{Weir}}, \bibinfo{journal}{Philosophical Transactions of the Royal Society A: Mathematical, Physical and Engineering Sciences} \textbf{\bibinfo{volume}{376}}, \bibinfo{pages}{20170126} (\bibinfo{year}{2018}).

\bibitem[{\citenamefont{Mazumdar and White}(2018)}]{mazumdar2018cosmic}
\bibinfo{author}{\bibfnamefont{A.}~\bibnamefont{Mazumdar}} \bibnamefont{and} \bibinfo{author}{\bibfnamefont{G.}~\bibnamefont{White}}, \bibinfo{journal}{arXiv preprint arXiv:1811.01948}  (\bibinfo{year}{2018}).

\bibitem[{\citenamefont{Bertone et~al.}(2020)\citenamefont{Bertone, Croon, Amin, Boddy, Kavanagh, Mack, Natarajan, Opferkuch, Schutz, Takhistov et~al.}}]{bertone2020gravitational}
\bibinfo{author}{\bibfnamefont{G.}~\bibnamefont{Bertone}}, \bibinfo{author}{\bibfnamefont{D.}~\bibnamefont{Croon}}, \bibinfo{author}{\bibfnamefont{M.}~\bibnamefont{Amin}}, \bibinfo{author}{\bibfnamefont{K.~K.} \bibnamefont{Boddy}}, \bibinfo{author}{\bibfnamefont{B.}~\bibnamefont{Kavanagh}}, \bibinfo{author}{\bibfnamefont{K.~J.} \bibnamefont{Mack}}, \bibinfo{author}{\bibfnamefont{P.}~\bibnamefont{Natarajan}}, \bibinfo{author}{\bibfnamefont{T.}~\bibnamefont{Opferkuch}}, \bibinfo{author}{\bibfnamefont{K.}~\bibnamefont{Schutz}}, \bibinfo{author}{\bibfnamefont{V.}~\bibnamefont{Takhistov}}, \bibnamefont{et~al.}, \bibinfo{journal}{SciPost Physics Core} \textbf{\bibinfo{volume}{3}}, \bibinfo{pages}{007} (\bibinfo{year}{2020}).

\bibitem[{\citenamefont{Abbott et~al.}(2017)\citenamefont{Abbott, Abbott, Abbott, Abernathy, Acernese, Ackley, Adams, Adams, Addesso, Adhikari et~al.}}]{abbott2017upper}
\bibinfo{author}{\bibfnamefont{B.~P.} \bibnamefont{Abbott}}, \bibinfo{author}{\bibfnamefont{R.}~\bibnamefont{Abbott}}, \bibinfo{author}{\bibfnamefont{T.~D.} \bibnamefont{Abbott}}, \bibinfo{author}{\bibfnamefont{M.}~\bibnamefont{Abernathy}}, \bibinfo{author}{\bibfnamefont{F.}~\bibnamefont{Acernese}}, \bibinfo{author}{\bibfnamefont{K.}~\bibnamefont{Ackley}}, \bibinfo{author}{\bibfnamefont{C.}~\bibnamefont{Adams}}, \bibinfo{author}{\bibfnamefont{T.}~\bibnamefont{Adams}}, \bibinfo{author}{\bibfnamefont{P.}~\bibnamefont{Addesso}}, \bibinfo{author}{\bibfnamefont{R.}~\bibnamefont{Adhikari}}, \bibnamefont{et~al.}, \bibinfo{journal}{Physical review letters} \textbf{\bibinfo{volume}{118}}, \bibinfo{pages}{121101} (\bibinfo{year}{2017}).

\bibitem[{\citenamefont{Scientific et~al.}(2019)\citenamefont{Scientific, Abbott, Abbott, Abbott, Abraham, Acernese, Ackley, Adams, Adya, Affeldt et~al.}}]{scientific2019search}
\bibinfo{author}{\bibfnamefont{L.}~\bibnamefont{Scientific}}, \bibinfo{author}{\bibfnamefont{B.}~\bibnamefont{Abbott}}, \bibinfo{author}{\bibfnamefont{R.}~\bibnamefont{Abbott}}, \bibinfo{author}{\bibfnamefont{T.}~\bibnamefont{Abbott}}, \bibinfo{author}{\bibfnamefont{S.}~\bibnamefont{Abraham}}, \bibinfo{author}{\bibfnamefont{F.}~\bibnamefont{Acernese}}, \bibinfo{author}{\bibfnamefont{K.}~\bibnamefont{Ackley}}, \bibinfo{author}{\bibfnamefont{C.}~\bibnamefont{Adams}}, \bibinfo{author}{\bibfnamefont{V.}~\bibnamefont{Adya}}, \bibinfo{author}{\bibfnamefont{C.}~\bibnamefont{Affeldt}}, \bibnamefont{et~al.}, \bibinfo{journal}{Physical Review D} \textbf{\bibinfo{volume}{100}}, \bibinfo{pages}{061101} (\bibinfo{year}{2019}).

\bibitem[{\citenamefont{Amaro-Seoane et~al.}(2017)\citenamefont{Amaro-Seoane, Audley, Babak, Baker, Barausse, Bender, Berti, Binetruy, Born, Bortoluzzi et~al.}}]{amaro2017laser}
\bibinfo{author}{\bibfnamefont{P.}~\bibnamefont{Amaro-Seoane}}, \bibinfo{author}{\bibfnamefont{H.}~\bibnamefont{Audley}}, \bibinfo{author}{\bibfnamefont{S.}~\bibnamefont{Babak}}, \bibinfo{author}{\bibfnamefont{J.}~\bibnamefont{Baker}}, \bibinfo{author}{\bibfnamefont{E.}~\bibnamefont{Barausse}}, \bibinfo{author}{\bibfnamefont{P.}~\bibnamefont{Bender}}, \bibinfo{author}{\bibfnamefont{E.}~\bibnamefont{Berti}}, \bibinfo{author}{\bibfnamefont{P.}~\bibnamefont{Binetruy}}, \bibinfo{author}{\bibfnamefont{M.}~\bibnamefont{Born}}, \bibinfo{author}{\bibfnamefont{D.}~\bibnamefont{Bortoluzzi}}, \bibnamefont{et~al.}, \bibinfo{journal}{arXiv preprint arXiv:1702.00786}  (\bibinfo{year}{2017}).

\bibitem[{\citenamefont{Gong et~al.}(2015)\citenamefont{Gong, Lau, Xu, Amaro-Seoane, Bai, Bian, Cao, Chen, Chen, Ding et~al.}}]{gong2015descope}
\bibinfo{author}{\bibfnamefont{X.}~\bibnamefont{Gong}}, \bibinfo{author}{\bibfnamefont{Y.-K.} \bibnamefont{Lau}}, \bibinfo{author}{\bibfnamefont{S.}~\bibnamefont{Xu}}, \bibinfo{author}{\bibfnamefont{P.}~\bibnamefont{Amaro-Seoane}}, \bibinfo{author}{\bibfnamefont{S.}~\bibnamefont{Bai}}, \bibinfo{author}{\bibfnamefont{X.}~\bibnamefont{Bian}}, \bibinfo{author}{\bibfnamefont{Z.}~\bibnamefont{Cao}}, \bibinfo{author}{\bibfnamefont{G.}~\bibnamefont{Chen}}, \bibinfo{author}{\bibfnamefont{X.}~\bibnamefont{Chen}}, \bibinfo{author}{\bibfnamefont{Y.}~\bibnamefont{Ding}}, \bibnamefont{et~al.}, in \emph{\bibinfo{booktitle}{Journal of Physics: Conference Series}} (\bibinfo{organization}{IOP Publishing}, \bibinfo{year}{2015}), vol. \bibinfo{volume}{610}, p. \bibinfo{pages}{012011}.

\bibitem[{\citenamefont{Luo et~al.}(2016)\citenamefont{Luo, Chen, Duan, Gong, Hu, Ji, Liu, Mei, Milyukov, Sazhin et~al.}}]{luo2016tianqin}
\bibinfo{author}{\bibfnamefont{J.}~\bibnamefont{Luo}}, \bibinfo{author}{\bibfnamefont{L.-S.} \bibnamefont{Chen}}, \bibinfo{author}{\bibfnamefont{H.-Z.} \bibnamefont{Duan}}, \bibinfo{author}{\bibfnamefont{Y.-G.} \bibnamefont{Gong}}, \bibinfo{author}{\bibfnamefont{S.}~\bibnamefont{Hu}}, \bibinfo{author}{\bibfnamefont{J.}~\bibnamefont{Ji}}, \bibinfo{author}{\bibfnamefont{Q.}~\bibnamefont{Liu}}, \bibinfo{author}{\bibfnamefont{J.}~\bibnamefont{Mei}}, \bibinfo{author}{\bibfnamefont{V.}~\bibnamefont{Milyukov}}, \bibinfo{author}{\bibfnamefont{M.}~\bibnamefont{Sazhin}}, \bibnamefont{et~al.}, \bibinfo{journal}{Classical and Quantum Gravity} \textbf{\bibinfo{volume}{33}}, \bibinfo{pages}{035010} (\bibinfo{year}{2016}).

\bibitem[{\citenamefont{Athron et~al.}(2023{\natexlab{a}})\citenamefont{Athron, Bal\'azs, Fowlie, Morris, and Wu}}]{Athron:2023xlk}
\bibinfo{author}{\bibfnamefont{P.}~\bibnamefont{Athron}}, \bibinfo{author}{\bibfnamefont{C.}~\bibnamefont{Bal\'azs}}, \bibinfo{author}{\bibfnamefont{A.}~\bibnamefont{Fowlie}}, \bibinfo{author}{\bibfnamefont{L.}~\bibnamefont{Morris}}, \bibnamefont{and} \bibinfo{author}{\bibfnamefont{L.}~\bibnamefont{Wu}} (\bibinfo{year}{2023}{\natexlab{a}}), \eprint{2305.02357}.

\bibitem[{\citenamefont{Han et~al.}(2021)\citenamefont{Han, Wang, and Zhang}}]{han2021dark}
\bibinfo{author}{\bibfnamefont{X.-F.} \bibnamefont{Han}}, \bibinfo{author}{\bibfnamefont{L.}~\bibnamefont{Wang}}, \bibnamefont{and} \bibinfo{author}{\bibfnamefont{Y.}~\bibnamefont{Zhang}}, \bibinfo{journal}{Physical Review D} \textbf{\bibinfo{volume}{103}}, \bibinfo{pages}{035012} (\bibinfo{year}{2021}).

\bibitem[{\citenamefont{Guo et~al.}(2021)\citenamefont{Guo, Sinha, Vagie, and White}}]{guo2021phase}
\bibinfo{author}{\bibfnamefont{H.-K.} \bibnamefont{Guo}}, \bibinfo{author}{\bibfnamefont{K.}~\bibnamefont{Sinha}}, \bibinfo{author}{\bibfnamefont{D.}~\bibnamefont{Vagie}}, \bibnamefont{and} \bibinfo{author}{\bibfnamefont{G.}~\bibnamefont{White}}, \bibinfo{journal}{Journal of Cosmology and Astroparticle Physics} \textbf{\bibinfo{volume}{2021}}, \bibinfo{pages}{001} (\bibinfo{year}{2021}).

\bibitem[{\citenamefont{Yang et~al.}(2023)\citenamefont{Yang, Zhou, and Bian}}]{yang2023gravitational}
\bibinfo{author}{\bibfnamefont{J.}~\bibnamefont{Yang}}, \bibinfo{author}{\bibfnamefont{R.}~\bibnamefont{Zhou}}, \bibnamefont{and} \bibinfo{author}{\bibfnamefont{L.}~\bibnamefont{Bian}}, \bibinfo{journal}{Physics Letters B} \textbf{\bibinfo{volume}{839}}, \bibinfo{pages}{137822} (\bibinfo{year}{2023}).

\bibitem[{\citenamefont{Xiao et~al.}(2023)\citenamefont{Xiao, Yang, and Zhang}}]{Xiao:2023dbb}
\bibinfo{author}{\bibfnamefont{Y.}~\bibnamefont{Xiao}}, \bibinfo{author}{\bibfnamefont{J.~M.} \bibnamefont{Yang}}, \bibnamefont{and} \bibinfo{author}{\bibfnamefont{Y.}~\bibnamefont{Zhang}} (\bibinfo{year}{2023}), \eprint{2307.01072}.

\bibitem[{\citenamefont{Bal\'azs et~al.}(2023)\citenamefont{Bal\'azs, Xiao, Yang, and Zhang}}]{Balazs:2023kuk}
\bibinfo{author}{\bibfnamefont{C.}~\bibnamefont{Bal\'azs}}, \bibinfo{author}{\bibfnamefont{Y.}~\bibnamefont{Xiao}}, \bibinfo{author}{\bibfnamefont{J.~M.} \bibnamefont{Yang}}, \bibnamefont{and} \bibinfo{author}{\bibfnamefont{Y.}~\bibnamefont{Zhang}} (\bibinfo{year}{2023}), \eprint{2301.09283}.

\bibitem[{\citenamefont{Vaskonen}(2017)}]{vaskonen2017electroweak}
\bibinfo{author}{\bibfnamefont{V.}~\bibnamefont{Vaskonen}}, \bibinfo{journal}{Physical Review D} \textbf{\bibinfo{volume}{95}}, \bibinfo{pages}{123515} (\bibinfo{year}{2017}).

\bibitem[{\citenamefont{Beniwal et~al.}(2019)\citenamefont{Beniwal, Lewicki, White, and Williams}}]{beniwal2019gravitational}
\bibinfo{author}{\bibfnamefont{A.}~\bibnamefont{Beniwal}}, \bibinfo{author}{\bibfnamefont{M.}~\bibnamefont{Lewicki}}, \bibinfo{author}{\bibfnamefont{M.}~\bibnamefont{White}}, \bibnamefont{and} \bibinfo{author}{\bibfnamefont{A.~G.} \bibnamefont{Williams}}, \bibinfo{journal}{Journal of High Energy Physics} \textbf{\bibinfo{volume}{2019}}, \bibinfo{pages}{1} (\bibinfo{year}{2019}).

\bibitem[{\citenamefont{Kang et~al.}(2018)\citenamefont{Kang, Ko, and Matsui}}]{kang2018strong}
\bibinfo{author}{\bibfnamefont{Z.}~\bibnamefont{Kang}}, \bibinfo{author}{\bibfnamefont{P.}~\bibnamefont{Ko}}, \bibnamefont{and} \bibinfo{author}{\bibfnamefont{T.}~\bibnamefont{Matsui}}, \bibinfo{journal}{Journal of High Energy Physics} \textbf{\bibinfo{volume}{2018}}, \bibinfo{pages}{1} (\bibinfo{year}{2018}).

\bibitem[{\citenamefont{Chala et~al.}(2018)\citenamefont{Chala, Krause, and Nardini}}]{chala2018signals}
\bibinfo{author}{\bibfnamefont{M.}~\bibnamefont{Chala}}, \bibinfo{author}{\bibfnamefont{C.}~\bibnamefont{Krause}}, \bibnamefont{and} \bibinfo{author}{\bibfnamefont{G.}~\bibnamefont{Nardini}}, \bibinfo{journal}{Journal of High Energy Physics} \textbf{\bibinfo{volume}{2018}}, \bibinfo{pages}{1} (\bibinfo{year}{2018}).

\bibitem[{\citenamefont{Alves et~al.}(2020)\citenamefont{Alves, Gon{\c{c}}alves, Ghosh, Guo, and Sinha}}]{alves2020di}
\bibinfo{author}{\bibfnamefont{A.}~\bibnamefont{Alves}}, \bibinfo{author}{\bibfnamefont{D.}~\bibnamefont{Gon{\c{c}}alves}}, \bibinfo{author}{\bibfnamefont{T.}~\bibnamefont{Ghosh}}, \bibinfo{author}{\bibfnamefont{H.-K.} \bibnamefont{Guo}}, \bibnamefont{and} \bibinfo{author}{\bibfnamefont{K.}~\bibnamefont{Sinha}}, \bibinfo{journal}{Journal of High Energy Physics} \textbf{\bibinfo{volume}{2020}}, \bibinfo{pages}{1} (\bibinfo{year}{2020}).

\bibitem[{\citenamefont{Alves et~al.}(2019)\citenamefont{Alves, Ghosh, Guo, Sinha, and Vagie}}]{alves2019collider}
\bibinfo{author}{\bibfnamefont{A.}~\bibnamefont{Alves}}, \bibinfo{author}{\bibfnamefont{T.}~\bibnamefont{Ghosh}}, \bibinfo{author}{\bibfnamefont{H.-K.} \bibnamefont{Guo}}, \bibinfo{author}{\bibfnamefont{K.}~\bibnamefont{Sinha}}, \bibnamefont{and} \bibinfo{author}{\bibfnamefont{D.}~\bibnamefont{Vagie}}, \bibinfo{journal}{Journal of High Energy Physics} \textbf{\bibinfo{volume}{2019}}, \bibinfo{pages}{1} (\bibinfo{year}{2019}).

\bibitem[{\citenamefont{Chao et~al.}(2017)\citenamefont{Chao, Guo, and Shu}}]{chao2017gravitational}
\bibinfo{author}{\bibfnamefont{W.}~\bibnamefont{Chao}}, \bibinfo{author}{\bibfnamefont{H.-K.} \bibnamefont{Guo}}, \bibnamefont{and} \bibinfo{author}{\bibfnamefont{J.}~\bibnamefont{Shu}}, \bibinfo{journal}{Journal of Cosmology and Astroparticle Physics} \textbf{\bibinfo{volume}{2017}}, \bibinfo{pages}{009} (\bibinfo{year}{2017}).

\bibitem[{\citenamefont{Kosowsky et~al.}(1992)\citenamefont{Kosowsky, Turner, and Watkins}}]{kosowsky1992gravitational}
\bibinfo{author}{\bibfnamefont{A.}~\bibnamefont{Kosowsky}}, \bibinfo{author}{\bibfnamefont{M.~S.} \bibnamefont{Turner}}, \bibnamefont{and} \bibinfo{author}{\bibfnamefont{R.}~\bibnamefont{Watkins}}, \bibinfo{journal}{Physical Review D} \textbf{\bibinfo{volume}{45}}, \bibinfo{pages}{4514} (\bibinfo{year}{1992}).

\bibitem[{\citenamefont{Kosowsky and Turner}(1993)}]{kosowsky1993gravitational}
\bibinfo{author}{\bibfnamefont{A.}~\bibnamefont{Kosowsky}} \bibnamefont{and} \bibinfo{author}{\bibfnamefont{M.~S.} \bibnamefont{Turner}}, \bibinfo{journal}{Physical Review D} \textbf{\bibinfo{volume}{47}}, \bibinfo{pages}{4372} (\bibinfo{year}{1993}).

\bibitem[{\citenamefont{Jinno and Takimoto}(2017)}]{jinno2017gravitational}
\bibinfo{author}{\bibfnamefont{R.}~\bibnamefont{Jinno}} \bibnamefont{and} \bibinfo{author}{\bibfnamefont{M.}~\bibnamefont{Takimoto}}, \bibinfo{journal}{Physical Review D} \textbf{\bibinfo{volume}{95}}, \bibinfo{pages}{024009} (\bibinfo{year}{2017}).

\bibitem[{\citenamefont{Hindmarsh et~al.}(2015)\citenamefont{Hindmarsh, Huber, Rummukainen, and Weir}}]{hindmarsh2015numerical}
\bibinfo{author}{\bibfnamefont{M.}~\bibnamefont{Hindmarsh}}, \bibinfo{author}{\bibfnamefont{S.~J.} \bibnamefont{Huber}}, \bibinfo{author}{\bibfnamefont{K.}~\bibnamefont{Rummukainen}}, \bibnamefont{and} \bibinfo{author}{\bibfnamefont{D.~J.} \bibnamefont{Weir}}, \bibinfo{journal}{Physical Review D} \textbf{\bibinfo{volume}{92}}, \bibinfo{pages}{123009} (\bibinfo{year}{2015}).

\bibitem[{\citenamefont{Hindmarsh et~al.}(2017)\citenamefont{Hindmarsh, Huber, Rummukainen, and Weir}}]{hindmarsh2017shape}
\bibinfo{author}{\bibfnamefont{M.}~\bibnamefont{Hindmarsh}}, \bibinfo{author}{\bibfnamefont{S.~J.} \bibnamefont{Huber}}, \bibinfo{author}{\bibfnamefont{K.}~\bibnamefont{Rummukainen}}, \bibnamefont{and} \bibinfo{author}{\bibfnamefont{D.~J.} \bibnamefont{Weir}}, \bibinfo{journal}{Physical Review D} \textbf{\bibinfo{volume}{96}}, \bibinfo{pages}{103520} (\bibinfo{year}{2017}).

\bibitem[{\citenamefont{Hindmarsh et~al.}(2014)\citenamefont{Hindmarsh, Huber, Rummukainen, and Weir}}]{hindmarsh2014gravitational}
\bibinfo{author}{\bibfnamefont{M.}~\bibnamefont{Hindmarsh}}, \bibinfo{author}{\bibfnamefont{S.~J.} \bibnamefont{Huber}}, \bibinfo{author}{\bibfnamefont{K.}~\bibnamefont{Rummukainen}}, \bibnamefont{and} \bibinfo{author}{\bibfnamefont{D.~J.} \bibnamefont{Weir}}, \bibinfo{journal}{Physical Review Letters} \textbf{\bibinfo{volume}{112}}, \bibinfo{pages}{041301} (\bibinfo{year}{2014}).

\bibitem[{\citenamefont{Hindmarsh}(2018)}]{hindmarsh2018sound}
\bibinfo{author}{\bibfnamefont{M.}~\bibnamefont{Hindmarsh}}, \bibinfo{journal}{Physical Review Letters} \textbf{\bibinfo{volume}{120}}, \bibinfo{pages}{071301} (\bibinfo{year}{2018}).

\bibitem[{\citenamefont{Hindmarsh and Hijazi}(2019)}]{hindmarsh2019gravitational}
\bibinfo{author}{\bibfnamefont{M.}~\bibnamefont{Hindmarsh}} \bibnamefont{and} \bibinfo{author}{\bibfnamefont{M.}~\bibnamefont{Hijazi}}, \bibinfo{journal}{Journal of Cosmology and Astroparticle Physics} \textbf{\bibinfo{volume}{2019}}, \bibinfo{pages}{062} (\bibinfo{year}{2019}).

\bibitem[{\citenamefont{Jinno and Takimoto}(2019)}]{jinno2019gravitational}
\bibinfo{author}{\bibfnamefont{R.}~\bibnamefont{Jinno}} \bibnamefont{and} \bibinfo{author}{\bibfnamefont{M.}~\bibnamefont{Takimoto}}, \bibinfo{journal}{Journal of Cosmology and Astroparticle Physics} \textbf{\bibinfo{volume}{2019}}, \bibinfo{pages}{060} (\bibinfo{year}{2019}).

\bibitem[{\citenamefont{Konstandin}(2018)}]{konstandin2018gravitational}
\bibinfo{author}{\bibfnamefont{T.}~\bibnamefont{Konstandin}}, \bibinfo{journal}{Journal of Cosmology and Astroparticle Physics} \textbf{\bibinfo{volume}{2018}}, \bibinfo{pages}{047} (\bibinfo{year}{2018}).

\bibitem[{\citenamefont{Cai et~al.}(2023{\natexlab{a}})\citenamefont{Cai, Wang, and Yuwen}}]{cai2023hydrodynamic}
\bibinfo{author}{\bibfnamefont{R.-G.} \bibnamefont{Cai}}, \bibinfo{author}{\bibfnamefont{S.-J.} \bibnamefont{Wang}}, \bibnamefont{and} \bibinfo{author}{\bibfnamefont{Z.-Y.} \bibnamefont{Yuwen}}, \bibinfo{journal}{arXiv preprint arXiv:2305.00074}  (\bibinfo{year}{2023}{\natexlab{a}}).

\bibitem[{\citenamefont{Kosowsky et~al.}(2002)\citenamefont{Kosowsky, Mack, and Kahniashvili}}]{kosowsky2002gravitational}
\bibinfo{author}{\bibfnamefont{A.}~\bibnamefont{Kosowsky}}, \bibinfo{author}{\bibfnamefont{A.}~\bibnamefont{Mack}}, \bibnamefont{and} \bibinfo{author}{\bibfnamefont{T.}~\bibnamefont{Kahniashvili}}, \bibinfo{journal}{Physical Review D} \textbf{\bibinfo{volume}{66}}, \bibinfo{pages}{024030} (\bibinfo{year}{2002}).

\bibitem[{\citenamefont{Dolgov et~al.}(2002)\citenamefont{Dolgov, Grasso, and Nicolis}}]{dolgov2002relic}
\bibinfo{author}{\bibfnamefont{A.~D.} \bibnamefont{Dolgov}}, \bibinfo{author}{\bibfnamefont{D.}~\bibnamefont{Grasso}}, \bibnamefont{and} \bibinfo{author}{\bibfnamefont{A.}~\bibnamefont{Nicolis}}, \bibinfo{journal}{Physical Review D} \textbf{\bibinfo{volume}{66}}, \bibinfo{pages}{103505} (\bibinfo{year}{2002}).

\bibitem[{\citenamefont{Caprini et~al.}(2009)\citenamefont{Caprini, Durrer, and Servant}}]{Caprini_2009}
\bibinfo{author}{\bibfnamefont{C.}~\bibnamefont{Caprini}}, \bibinfo{author}{\bibfnamefont{R.}~\bibnamefont{Durrer}}, \bibnamefont{and} \bibinfo{author}{\bibfnamefont{G.}~\bibnamefont{Servant}}, \bibinfo{journal}{Journal of Cosmology and Astroparticle Physics} \textbf{\bibinfo{volume}{2009}}, \bibinfo{pages}{024} (\bibinfo{year}{2009}).

\bibitem[{\citenamefont{Yang and Bian}(2022)}]{Yang:2021uid}
\bibinfo{author}{\bibfnamefont{J.}~\bibnamefont{Yang}} \bibnamefont{and} \bibinfo{author}{\bibfnamefont{L.}~\bibnamefont{Bian}}, \bibinfo{journal}{Phys. Rev. D} \textbf{\bibinfo{volume}{106}}, \bibinfo{pages}{023510} (\bibinfo{year}{2022}), \eprint{2102.01398}.

\bibitem[{\citenamefont{Di et~al.}(2021)\citenamefont{Di, Wang, Zhou, Bian, Cai, and Liu}}]{Di:2020kbw}
\bibinfo{author}{\bibfnamefont{Y.}~\bibnamefont{Di}}, \bibinfo{author}{\bibfnamefont{J.}~\bibnamefont{Wang}}, \bibinfo{author}{\bibfnamefont{R.}~\bibnamefont{Zhou}}, \bibinfo{author}{\bibfnamefont{L.}~\bibnamefont{Bian}}, \bibinfo{author}{\bibfnamefont{R.-G.} \bibnamefont{Cai}}, \bibnamefont{and} \bibinfo{author}{\bibfnamefont{J.}~\bibnamefont{Liu}}, \bibinfo{journal}{Phys. Rev. Lett.} \textbf{\bibinfo{volume}{126}}, \bibinfo{pages}{251102} (\bibinfo{year}{2021}), \eprint{2012.15625}.

\bibitem[{\citenamefont{Croon et~al.}(2021)\citenamefont{Croon, Gould, Schicho, Tenkanen, and White}}]{croon2021theoretical}
\bibinfo{author}{\bibfnamefont{D.}~\bibnamefont{Croon}}, \bibinfo{author}{\bibfnamefont{O.}~\bibnamefont{Gould}}, \bibinfo{author}{\bibfnamefont{P.}~\bibnamefont{Schicho}}, \bibinfo{author}{\bibfnamefont{T.~V.} \bibnamefont{Tenkanen}}, \bibnamefont{and} \bibinfo{author}{\bibfnamefont{G.}~\bibnamefont{White}}, \bibinfo{journal}{Journal of High Energy Physics} \textbf{\bibinfo{volume}{2021}}, \bibinfo{pages}{1} (\bibinfo{year}{2021}).

\bibitem[{\citenamefont{Athron et~al.}(2023{\natexlab{b}})\citenamefont{Athron, Balazs, Fowlie, Morris, White, and Zhang}}]{Athron:2022jyi}
\bibinfo{author}{\bibfnamefont{P.}~\bibnamefont{Athron}}, \bibinfo{author}{\bibfnamefont{C.}~\bibnamefont{Balazs}}, \bibinfo{author}{\bibfnamefont{A.}~\bibnamefont{Fowlie}}, \bibinfo{author}{\bibfnamefont{L.}~\bibnamefont{Morris}}, \bibinfo{author}{\bibfnamefont{G.}~\bibnamefont{White}}, \bibnamefont{and} \bibinfo{author}{\bibfnamefont{Y.}~\bibnamefont{Zhang}}, \bibinfo{journal}{JHEP} \textbf{\bibinfo{volume}{01}}, \bibinfo{pages}{050} (\bibinfo{year}{2023}{\natexlab{b}}), \eprint{2208.01319}.

\bibitem[{\citenamefont{Andreassen et~al.}(2017)\citenamefont{Andreassen, Farhi, Frost, and Schwartz}}]{andreassen2017precision}
\bibinfo{author}{\bibfnamefont{A.}~\bibnamefont{Andreassen}}, \bibinfo{author}{\bibfnamefont{D.}~\bibnamefont{Farhi}}, \bibinfo{author}{\bibfnamefont{W.}~\bibnamefont{Frost}}, \bibnamefont{and} \bibinfo{author}{\bibfnamefont{M.~D.} \bibnamefont{Schwartz}}, \bibinfo{journal}{Physical Review D} \textbf{\bibinfo{volume}{95}}, \bibinfo{pages}{085011} (\bibinfo{year}{2017}).

\bibitem[{\citenamefont{Dunne and Min}(2005)}]{dunne2005beyond}
\bibinfo{author}{\bibfnamefont{G.~V.} \bibnamefont{Dunne}} \bibnamefont{and} \bibinfo{author}{\bibfnamefont{H.}~\bibnamefont{Min}}, \bibinfo{journal}{Physical Review D} \textbf{\bibinfo{volume}{72}}, \bibinfo{pages}{125004} (\bibinfo{year}{2005}).

\bibitem[{\citenamefont{Ivanov et~al.}(2022)\citenamefont{Ivanov, Matteini, Nemev\v{s}ek, and Ubaldi}}]{Ivanov:2022osf}
\bibinfo{author}{\bibfnamefont{A.}~\bibnamefont{Ivanov}}, \bibinfo{author}{\bibfnamefont{M.}~\bibnamefont{Matteini}}, \bibinfo{author}{\bibfnamefont{M.}~\bibnamefont{Nemev\v{s}ek}}, \bibnamefont{and} \bibinfo{author}{\bibfnamefont{L.}~\bibnamefont{Ubaldi}}, \bibinfo{journal}{JHEP} \textbf{\bibinfo{volume}{03}}, \bibinfo{pages}{209} (\bibinfo{year}{2022}), \bibinfo{note}{[Erratum: JHEP 07, 085 (2022), Erratum: JHEP 11, 157 (2022)]}, \eprint{2202.04498}.

\bibitem[{\citenamefont{Ekstedt}(2022)}]{ekstedt2022higher}
\bibinfo{author}{\bibfnamefont{A.}~\bibnamefont{Ekstedt}}, \bibinfo{journal}{The European Physical Journal C} \textbf{\bibinfo{volume}{82}}, \bibinfo{pages}{173} (\bibinfo{year}{2022}).

\bibitem[{\citenamefont{Ekstedt et~al.}(2023)\citenamefont{Ekstedt, Gould, and Hirvonen}}]{Ekstedt:2023sqc}
\bibinfo{author}{\bibfnamefont{A.}~\bibnamefont{Ekstedt}}, \bibinfo{author}{\bibfnamefont{O.}~\bibnamefont{Gould}}, \bibnamefont{and} \bibinfo{author}{\bibfnamefont{J.}~\bibnamefont{Hirvonen}}, \bibinfo{journal}{JHEP} \textbf{\bibinfo{volume}{12}}, \bibinfo{pages}{056} (\bibinfo{year}{2023}), \eprint{2308.15652}.

\bibitem[{\citenamefont{Guth and Weinberg}(1981)}]{guth1981cosmological}
\bibinfo{author}{\bibfnamefont{A.~H.} \bibnamefont{Guth}} \bibnamefont{and} \bibinfo{author}{\bibfnamefont{E.~J.} \bibnamefont{Weinberg}}, \bibinfo{journal}{Physical Review D} \textbf{\bibinfo{volume}{23}}, \bibinfo{pages}{876} (\bibinfo{year}{1981}).

\bibitem[{\citenamefont{Enqvist et~al.}(1992)\citenamefont{Enqvist, Ignatius, Kajantie, and Rummukainen}}]{enqvist1992nucleation}
\bibinfo{author}{\bibfnamefont{K.}~\bibnamefont{Enqvist}}, \bibinfo{author}{\bibfnamefont{J.}~\bibnamefont{Ignatius}}, \bibinfo{author}{\bibfnamefont{K.}~\bibnamefont{Kajantie}}, \bibnamefont{and} \bibinfo{author}{\bibfnamefont{K.}~\bibnamefont{Rummukainen}}, \bibinfo{journal}{Physical Review D} \textbf{\bibinfo{volume}{45}}, \bibinfo{pages}{3415} (\bibinfo{year}{1992}).

\bibitem[{\citenamefont{Athron et~al.}(2023{\natexlab{c}})\citenamefont{Athron, Bal{\'a}zs, and Morris}}]{athron2023supercool}
\bibinfo{author}{\bibfnamefont{P.}~\bibnamefont{Athron}}, \bibinfo{author}{\bibfnamefont{C.}~\bibnamefont{Bal{\'a}zs}}, \bibnamefont{and} \bibinfo{author}{\bibfnamefont{L.}~\bibnamefont{Morris}}, \bibinfo{journal}{Journal of Cosmology and Astroparticle Physics} \textbf{\bibinfo{volume}{2023}}, \bibinfo{pages}{006} (\bibinfo{year}{2023}{\natexlab{c}}).

\bibitem[{\citenamefont{Cai et~al.}(2017)\citenamefont{Cai, Sasaki, and Wang}}]{cai2017gravitational}
\bibinfo{author}{\bibfnamefont{R.-G.} \bibnamefont{Cai}}, \bibinfo{author}{\bibfnamefont{M.}~\bibnamefont{Sasaki}}, \bibnamefont{and} \bibinfo{author}{\bibfnamefont{S.-J.} \bibnamefont{Wang}}, \bibinfo{journal}{Journal of Cosmology and Astroparticle Physics} \textbf{\bibinfo{volume}{2017}}, \bibinfo{pages}{004} (\bibinfo{year}{2017}).

\bibitem[{\citenamefont{Linde}(1983)}]{linde1983decay}
\bibinfo{author}{\bibfnamefont{A.~D.} \bibnamefont{Linde}}, \bibinfo{journal}{Nuclear Physics B} \textbf{\bibinfo{volume}{216}}, \bibinfo{pages}{421} (\bibinfo{year}{1983}).

\bibitem[{\citenamefont{Rubakov}(2009)}]{rubakov2009classical}
\bibinfo{author}{\bibfnamefont{V.}~\bibnamefont{Rubakov}}, in \emph{\bibinfo{booktitle}{Classical Theory of Gauge Fields}} (\bibinfo{publisher}{Princeton University Press}, \bibinfo{year}{2009}).

\bibitem[{\citenamefont{Ajmi and Hindmarsh}(2022)}]{Ajmi:2022nmq}
\bibinfo{author}{\bibfnamefont{M.~A.} \bibnamefont{Ajmi}} \bibnamefont{and} \bibinfo{author}{\bibfnamefont{M.}~\bibnamefont{Hindmarsh}}, \bibinfo{journal}{Phys. Rev. D} \textbf{\bibinfo{volume}{106}}, \bibinfo{pages}{023505} (\bibinfo{year}{2022}), \eprint{2205.04097}.

\bibitem[{\citenamefont{Imai et~al.}(1985)\citenamefont{Imai, Iri, and Murota}}]{imai1985voronoi}
\bibinfo{author}{\bibfnamefont{H.}~\bibnamefont{Imai}}, \bibinfo{author}{\bibfnamefont{M.}~\bibnamefont{Iri}}, \bibnamefont{and} \bibinfo{author}{\bibfnamefont{K.}~\bibnamefont{Murota}}, \bibinfo{journal}{SIAM Journal on Computing} \textbf{\bibinfo{volume}{14}}, \bibinfo{pages}{93} (\bibinfo{year}{1985}).

\bibitem[{\citenamefont{Hijazi}(2022)}]{hijazi2022numerical}
\bibinfo{author}{\bibfnamefont{M.}~\bibnamefont{Hijazi}}, \bibinfo{journal}{arXiv preprint arXiv:2208.10636}  (\bibinfo{year}{2022}).

\bibitem[{\citenamefont{Espinosa et~al.}(2010)\citenamefont{Espinosa, Konstandin, No, and Servant}}]{espinosa2010energy}
\bibinfo{author}{\bibfnamefont{J.~R.} \bibnamefont{Espinosa}}, \bibinfo{author}{\bibfnamefont{T.}~\bibnamefont{Konstandin}}, \bibinfo{author}{\bibfnamefont{J.~M.} \bibnamefont{No}}, \bibnamefont{and} \bibinfo{author}{\bibfnamefont{G.}~\bibnamefont{Servant}}, \bibinfo{journal}{Journal of Cosmology and Astroparticle Physics} \textbf{\bibinfo{volume}{2010}}, \bibinfo{pages}{028} (\bibinfo{year}{2010}).

\bibitem[{\citenamefont{Cai et~al.}(2023{\natexlab{b}})\citenamefont{Cai, Wang, and Yuwen}}]{Cai:2023guc}
\bibinfo{author}{\bibfnamefont{R.-G.} \bibnamefont{Cai}}, \bibinfo{author}{\bibfnamefont{S.-J.} \bibnamefont{Wang}}, \bibnamefont{and} \bibinfo{author}{\bibfnamefont{Z.-Y.} \bibnamefont{Yuwen}}, \bibinfo{journal}{Phys. Rev. D} \textbf{\bibinfo{volume}{108}}, \bibinfo{pages}{L021502} (\bibinfo{year}{2023}{\natexlab{b}}), \eprint{2305.00074}.

\bibitem[{\citenamefont{Gowling and Hindmarsh}(2021)}]{Gowling_2021}
\bibinfo{author}{\bibfnamefont{C.}~\bibnamefont{Gowling}} \bibnamefont{and} \bibinfo{author}{\bibfnamefont{M.}~\bibnamefont{Hindmarsh}}, \bibinfo{journal}{Journal of Cosmology and Astroparticle Physics} \textbf{\bibinfo{volume}{2021}}, \bibinfo{pages}{039} (\bibinfo{year}{2021}).

\bibitem[{\citenamefont{Cutting et~al.}(2020)\citenamefont{Cutting, Hindmarsh, and Weir}}]{Cutting:2019zws}
\bibinfo{author}{\bibfnamefont{D.}~\bibnamefont{Cutting}}, \bibinfo{author}{\bibfnamefont{M.}~\bibnamefont{Hindmarsh}}, \bibnamefont{and} \bibinfo{author}{\bibfnamefont{D.~J.} \bibnamefont{Weir}}, \bibinfo{journal}{Phys. Rev. Lett.} \textbf{\bibinfo{volume}{125}}, \bibinfo{pages}{021302} (\bibinfo{year}{2020}), \eprint{1906.00480}.

\bibitem[{\citenamefont{Adams}(1993)}]{adams1993general}
\bibinfo{author}{\bibfnamefont{F.~C.} \bibnamefont{Adams}}, \bibinfo{journal}{Physical Review D} \textbf{\bibinfo{volume}{48}}, \bibinfo{pages}{2800} (\bibinfo{year}{1993}).

\bibitem[{\citenamefont{Dine et~al.}(1992)\citenamefont{Dine, Leigh, Huet, Linde, and Linde}}]{Dine:1992wr}
\bibinfo{author}{\bibfnamefont{M.}~\bibnamefont{Dine}}, \bibinfo{author}{\bibfnamefont{R.~G.} \bibnamefont{Leigh}}, \bibinfo{author}{\bibfnamefont{P.~Y.} \bibnamefont{Huet}}, \bibinfo{author}{\bibfnamefont{A.~D.} \bibnamefont{Linde}}, \bibnamefont{and} \bibinfo{author}{\bibfnamefont{D.~A.} \bibnamefont{Linde}}, \bibinfo{journal}{Phys. Rev. D} \textbf{\bibinfo{volume}{46}}, \bibinfo{pages}{550} (\bibinfo{year}{1992}), \eprint{hep-ph/9203203}.

\bibitem[{\citenamefont{Lewicki et~al.}(2022)\citenamefont{Lewicki, Merchand, and Zych}}]{lewicki2022electroweak}
\bibinfo{author}{\bibfnamefont{M.}~\bibnamefont{Lewicki}}, \bibinfo{author}{\bibfnamefont{M.}~\bibnamefont{Merchand}}, \bibnamefont{and} \bibinfo{author}{\bibfnamefont{M.}~\bibnamefont{Zych}}, \bibinfo{journal}{Journal of High Energy Physics} \textbf{\bibinfo{volume}{2022}}, \bibinfo{pages}{1} (\bibinfo{year}{2022}).

\bibitem[{\citenamefont{Konstandin et~al.}(2014)\citenamefont{Konstandin, Nardini, and Rues}}]{konstandin2014boltzmann}
\bibinfo{author}{\bibfnamefont{T.}~\bibnamefont{Konstandin}}, \bibinfo{author}{\bibfnamefont{G.}~\bibnamefont{Nardini}}, \bibnamefont{and} \bibinfo{author}{\bibfnamefont{I.}~\bibnamefont{Rues}}, \bibinfo{journal}{Journal of Cosmology and Astroparticle Physics} \textbf{\bibinfo{volume}{2014}}, \bibinfo{pages}{028} (\bibinfo{year}{2014}).

\bibitem[{\citenamefont{Wang et~al.}(2020)\citenamefont{Wang, Huang, and Zhang}}]{wang2020bubble}
\bibinfo{author}{\bibfnamefont{X.}~\bibnamefont{Wang}}, \bibinfo{author}{\bibfnamefont{F.~P.} \bibnamefont{Huang}}, \bibnamefont{and} \bibinfo{author}{\bibfnamefont{X.}~\bibnamefont{Zhang}}, \bibinfo{journal}{arXiv preprint arXiv:2011.12903}  (\bibinfo{year}{2020}).

\bibitem[{\citenamefont{M{\'e}gevand and S{\'a}nchez}(2010)}]{megevand2010velocity}
\bibinfo{author}{\bibfnamefont{A.}~\bibnamefont{M{\'e}gevand}} \bibnamefont{and} \bibinfo{author}{\bibfnamefont{A.~D.} \bibnamefont{S{\'a}nchez}}, \bibinfo{journal}{Nuclear physics B} \textbf{\bibinfo{volume}{825}}, \bibinfo{pages}{151} (\bibinfo{year}{2010}).

\bibitem[{\citenamefont{Jinno et~al.}(2019)\citenamefont{Jinno, Konstandin, and Takimoto}}]{jinno2019relativistic}
\bibinfo{author}{\bibfnamefont{R.}~\bibnamefont{Jinno}}, \bibinfo{author}{\bibfnamefont{T.}~\bibnamefont{Konstandin}}, \bibnamefont{and} \bibinfo{author}{\bibfnamefont{M.}~\bibnamefont{Takimoto}}, \bibinfo{journal}{Journal of Cosmology and Astroparticle Physics} \textbf{\bibinfo{volume}{2019}}, \bibinfo{pages}{035} (\bibinfo{year}{2019}).

\bibitem[{\citenamefont{Gould et~al.}(2021)\citenamefont{Gould, Sukuvaara, and Weir}}]{gould2021vacuum}
\bibinfo{author}{\bibfnamefont{O.}~\bibnamefont{Gould}}, \bibinfo{author}{\bibfnamefont{S.}~\bibnamefont{Sukuvaara}}, \bibnamefont{and} \bibinfo{author}{\bibfnamefont{D.}~\bibnamefont{Weir}}, \bibinfo{journal}{Physical Review D} \textbf{\bibinfo{volume}{104}}, \bibinfo{pages}{075039} (\bibinfo{year}{2021}).

\end{thebibliography}
\end{document}